\documentclass{aa}
\usepackage{graphicx}
\begin{document}
\newcommand{\kms}{km~s$^{-1}$}
\newcommand{\cm}{cm$^{-2}$}
\newcommand{\lya}{Lyman~$\alpha$}
\newcommand{\lyb}{Lyman~$\beta$}
\newcommand{\za}{$z_{\rm abs}$}
\newcommand{\ze}{$z_{\rm em}$}
\newcommand{\nhi}{$N$(H~I)}

\def\ltsima{$\; \buildrel < \over \sim \;$}
\def\simlt{\lower.5ex\hbox{\ltsima}}
\def\gtsima{$\; \buildrel > \over \sim \;$}
\def\simgt{\lower.5ex\hbox{\gtsima}}
\def\arcs{$''~$}
\def\arcm{$'~$}

   \title{The CORALS Survey I:  New Estimates of the Number Density 
   and Gas Content of Damped Lyman Alpha Systems 
   Free from Dust Bias.\thanks{The work presented here is based in part 
on data obtained with the ESO facilities on La Silla (EFOSC/3.6-m) and Paranal
(FORS1/UT1). }
}

   \titlerunning{CORALS I: DLA number density and $\Omega_{\rm DLA}$.}

   \author{S. L. Ellison\inst{1}
          \and
          L. Yan\inst{2}
	  \and
	  I. M. Hook\inst{3}
	\and
	M. Pettini\inst{4}
	\and
	J. V. Wall\inst{5}
	\and
	P. Shaver\inst{6}
          }

   \offprints{S. Ellison}

   \institute{European Southern Observatory, Casilla 19001, Santiago 19, 
	Chile\\
              \email{sellison@eso.org}
        \and
	SIRTF Science Center, Caltech, California, USA\\
	 \email{lyan@ipac.caltech.edu}
	\and
	Astrophysics: Department of Physics, Nuclear and Astrophysics 
        Laboratory, Keble Road, Oxford,  OX1 3RH, UK\\
	\email{ihook@gemini.edu}
	\and
	Institute of Astronomy, Madingley Rd., Cambridge, CB3 0HA, UK\\
	\email{pettini@ast.cam.ac.uk}
	\and
	Astrophysics: Department of Physics, Nuclear and Astrophysics 
        Laboratory, Keble Road, Oxford,  OX1 3RH, UK\\
	\email{jvw@astro.ox.ac.uk}
	\and
	European Southern Observatory, Karl-Schwarzschild-Str. 2, D-85748
        Garching bei Munchen, Germany
	\email{pshaver@eso.org}
}

   \date{Received/accepted}

   \abstract{
We present the first results from the Complete Optical and Radio
Absorption Line System (CORALS) survey. We have compiled a
homogeneous sample of radio-selected QSOs from the Parkes
Catalogue and searched for damped Lyman alpha systems (DLAs) towards
every target, irrespective of its optical magnitude. This
approach circumvents selection effects -- particularly from
intervening dust -- which have long been suspected to affect DLA
surveys in optically-selected, magnitude-limited QSO samples.
The CORALS data set consists of 66 $z_{\rm em} \geq 2.2$ QSOs in
which 22 DLAs with absorption redshifts $1.8 \leq z_{\rm abs}
\leq z_{\rm em}$ have been identified over a total redshift interval
$\Delta z = 55.46$\,. Three of the DLAs are classified as
`associated' systems with $z_{\rm abs} \sim z_{\rm em}$; of the
19 intervening DLAs, 17 are new discoveries.  In this first paper
of the CORALS series we describe the sample, present intermediate
resolution spectroscopy and determine the population statistics
of DLAs.  We deduce a value of the neutral gas mass density
traced by DLAs (expressed as a fraction of the closure density)
$\log \Omega_{\rm DLA} h = -2.59^{+0.17}_{-0.24}$, and a number
density of DLAs per unit redshift $n(z) = 0.31^{+0.09}_{-0.08}$,
both at a mean redshift $\langle z \rangle = 2.37$. Both values
are only marginally higher than those measured in optically
selected samples of QSOs.  Taking into account the errors, we
conclude that dust-induced bias in previous surveys may have led
to an underestimate of these quantities by at most a factor of
two. While $n(z)$ is greater in fainter ($B > 20$) QSOs, the
effect is only at the $\sim 1 \sigma$ level and we have not
uncovered a previously unrecognised population of high column
density ($N$(H~I)$ > 10^{21}$\,cm$^{-2}$) DLAs in front of faint
QSOs. These conclusions are tentative because of the
limited size of our data set; in particular the distribution of
column densities is poorly sampled at the high end where a much
larger survey of radio-selected QSOs is required the improve the
statistics.
   \keywords{Quasars: absorption lines -- dust, extinction -- 
galaxies: evolution}
   }

   \maketitle
%

\section{Introduction}

The presence of intervening gas in the sightline to a high
redshift QSO will cause absorption lines to be superimposed
on the quasar's continuum.  This relatively simple physical
process allows us to study in detail the properties of gas
clouds in front of QSOs, whether in the intergalactic medium
or in the galaxies giving rise to high column density
damped Lyman alpha Systems (DLAs).  DLAs are one of the key
components of the universe at high redshift because,
although relatively rare, they account for most of the
neutral gas available for star formation. 

Measuring the redshift evolution of $\Omega_{\rm DLA}$ (the total
amount of neutral gas traced by DLAs expressed as a fraction of the
closure density) has traditionally been seen as a tool for
probing the history of assembling galaxies and measuring the rate
at which they convert gas into stars (e.g. Lanzetta, Wolfe, \&
Turnshek 1995; Pei \& Fall 1995; Storrie-Lombardi, McMahon \&
Irwin 1996; Storrie-Lombardi \& Wolfe 2000; Rao \& Turnshek
2000). Similarly, chemical abundances in DLAs have been used to
trace the metallicity evolution of galaxies over most of the
Hubble time (e.g. Pettini et al. 1997; Pettini et al. 1999;
Prochaska \& Wolfe 1999, 2000). However, the interpretation 
of both DLA abundances and $\Omega_{\rm DLA}$ as indicators of 
galaxy evolution 
is pivotal upon the assumption that DLAs are
representative of the bulk of `normal' galaxies at each redshift
sampled.  Therefore, it is rather surprising that, at least for
current samples of DLAs, no strong evolution of either
$\Omega_{\rm DLA}$ or metallicity ($Z_{\rm DLA}$) has been revealed
by the above studies from $z \sim 3.5$ down to $z \sim 0.5$\,.
One possible explanation is that DLAs are representative of only
a particular evolutionary stage in a galaxy's lifetime. 
Alternatively, current DLA
samples may be incomplete due to a selection bias 
In this paper, we specifically investigate the possibility that
dust extinction may have produced incomplete samples of
DLAs in previous surveys based on optically-selected QSOs.

DLAs appear to remain relatively metal-poor 
($Z_{\rm DLA} \approx \frac{1}{10} - \frac{1}{30} Z_{\odot}$) 
from $z \sim 3.5$ to $z \sim 0.5$
(Pettini et al. 1999; Prochaska \& Wolfe 2000). 
However, it has been noted by several authors that the
absorbers with the highest values of hydrogen column density
are all of low metallicity and dust content
(e.g. Pettini et al. 1997; Boiss\'{e} et al. 1998; 
Pettini et al. 1999; Prantzos and Boissier 2000; Savaglio 2001). Whilst
relatively metal-rich damped systems do exist, they tend to
have values of \nhi\ towards the lower end of the range
considered for DLAs, $N$(H~I)$\geq 2 \times
10^{20}$~cm$^{-2}$. Boiss\'{e} et al. (1998) and Prantzos \&
Boissier (2000) in particular have pointed out an apparent
anti-correlation between \nhi\ and metallicity. 
Since the census of metal abundances (using the column-density weighted
average) is dominated by the highest \nhi\ absorbers, such an 
anti-correlation could
explain the lack of metallicity evolution in the present
samples of DLAs. This anti-correlation between \nhi\ and metallicity
may reflect a bias against high column density,
metal-rich DLAs with sufficient dust to obscure background
QSOs from the eyes of magnitude-limited optical surveys.

Further evidence that optical magnitude-limited QSO samples may be
incomplete due to dust bias comes from consideration of the
spectral energy distribution of quasars with and without
DLAs.  Fall, Pei, \& McMahon (1989) and Pei, Fall, \&
Bechtold (1991) found evidence that QSOs with DLAs have
systematically steeper continuum slopes than quasars with no
damped absorber.  A similar conclusion was reached by
Carilli et al. (1998) who found that a high fraction of
their `red' QSO sample had associated 21cm absorbers
compared with QSOs selected solely for the presence of Mg~II
absorbers.  By modelling the extinction effect of dust, Pei
\& Fall (1995) estimated that at $z = 3$ between 27\% and
44\% of QSOs and 23\%--38\% of DLAs may be missing from
$B$-band selected quasar samples.

The objective of the Complete Optical and Radio Absorption
Line System (CORALS) survey presented here is simple and
straightforward; compile a homogeneous sample of
radio-selected QSOs and obtain medium resolution spectra of
every target, regardless of optical magnitude.  In this way
it should be possible to determine quantitatively the
severity of the dust bias implied by the observations
reviewed above. The present paper is the first in the CORALS
series. Here we define the sample (\S\ref{def_sec}),
describe the observations and present the QSO spectra
(\S\ref{obs_sec}), identify the DLAs (\S\ref{dla_sec}), and
calculate their number density and $\Omega_{\rm DLA}$
(\S\ref{stats_sec}).

We adopt an $\Omega_M = 1.0, \Omega_{\Lambda} = 0$ cosmology and
use $H_0 = 65$ \kms Mpc$^{-1}$ throughout the paper.

\section{Previous DLA Surveys}\label{prev_sec}

\begin{table*}
\begin{tabular}{lcccccc}\hline \hline
 & & & & & & \\
Survey reference & Abbreviation & No. QSOs & Supplemented &
Limiting Mag. & Telescope & Resolution \\
 & & & with & (Band) & & \\ \hline
 & & & & & & \\
Wolfe et al. (1986) & WTSC86 & 68 & ... & 18.5($V$) & Lick & 10 \AA \\
 & & & & & & \\
Sargent, Steidel \&  & SSB89 & 53 & ... & N/A$^{a}$
& Hale 5-m & 4 \AA  \\
Boksenberg (1989) & & & & & & \\
 & & & & & & \\
Lanzetta et al. (1991) & LAN91 & 57 & SSB89 & Literature & Hale
5-m and Las & 4 -- 6 \AA \\
 & & & & search & Campanas 2.5-m & \\
 & & & & & & \\
Lanzetta, Wolfe \&  & LWT95 & 260 & LAN91 &
20($V$)$^{b}$ & IUE & 7 -- 10 \AA \\
Turnshek (1995) & & & & & & \\
 & & & & & & \\
Wolfe et al. (1995) & WLFC95 & 228 & LWT95 & 18.75($B$) & MMT and Las 
& 6 -- 10 \AA \\
 & & & & & Campanas 2.5-m & \\ 
 & & & & & & \\
Storrie-Lombardi & SLW00 & 40 & Many incl. WTSC86, & 19.5($R$)$^{b}$
 & Keck, AAT & 2 \AA \\
  \& Wolfe (2000)  &  & & LWT95, WLFC95 & & Lick  & \\
& & & & & & \\
P\'{e}roux et al. (2001a) & PER01 & 66 & SLW00 & 20.5($R$)$^{b}$ & 
4-m CTIO & 5 \AA \\
 & & & & & WHT & \\ \hline
\end{tabular}
\caption{\label{past_surveys} Summary of past surveys and their sample
definitions.  Column 3 tabulates the number of QSOs new to each
survey.  Therefore, if a particular survey has been supplemented with
a previous sample, the total number of QSOs is the sum of the
individual entries in column 3.
Notes: 
$^{a}$ -- No broad-band limit is
quoted for SSB89, only a derived 
apparent magnitude from the continuum flux at $\lambda_0 = 1450$
\AA. 
$^{b}$ -- This is the faintest magnitude in this 
sample, but it does not indicate a completeness limit.}
\end{table*}

It is useful at this point to review some of the previous surveys for
damped \lya\ systems and summarise their sample
definitions and major findings. Relevant data are collected
in Table \ref{past_surveys}.

The first major survey for DLAs was conducted by Wolfe et
al. (1986) (WTSC86) from the relatively bright Lick QSO
sample. The strategy adopted by those authors was to obtain
low resolution spectra for a large number of QSOs in order
to identify absorption systems above a certain equivalent
width limit (in this case $W_0 \ge 5$ \AA).  These candidate
DLAs were then followed-up with 2 \AA\ resolution
spectroscopy revealing that approximately 50\% were indeed
damped, the remainder being blends of lower column density
lines. This strategy of pre-selecting DLA candidates from
low resolution spectra, based on rest-frame equivalent
width, has come to characterise surveys in subsequent years.
The approach allows a large sample of QSOs to be
considered, maximising efficient use of observing time.  The
WTSC86 survey represents a landmark in absorption line
studies, being the first major compilation of spectra and
line lists for DLAs.  One of the main results was that the
incidence of DLAs was found to be greater than expected if
the absorption were due to galaxies with the same cross
section for H~I absorption as present-day spirals (Wolfe
1988). In addition, the procedures for identifying DLAs were
established and the $N$(H~I) definition of a damped system
set by this work; $N$(H~I)$\ge 2\times 10^{20}$ \cm\ has
been upheld in all subsequent studies. 

A survey of similar size to the WTSC86 sample was conducted
a few years later by Sargent, Steidel \& Boksenberg (1989)
(SSB89), although the initial focus of this work was on
Lyman limit systems (LLSs).  All of the quasars in the SSB89
sample were observed at a resolution of 4 \AA, considerably
higher than the first pass made by Wolfe et al. (1986).
However, SSB89 estimate that their DLA sample is probably
only complete for systems with $N$(H~I)$\ge 10^{21}$ \cm,
although several lower column density lines were identified.
Combining the SSB89 dataset with a similar number of QSOs
found in the literature, Lanzetta et al. (1991) adopted the
strategy of WTSC86 in making a 5 \AA\ equivalent width cut
to select 89 DLA candidates from their sample of 101 QSOs.  

From follow-up spectroscopy at 2 -- 3 \AA\ resolution of
these candidates, Lanzetta et al. (1991) produced the first
thorough statistical analysis of damped system number
density, the mass density of neutral gas in DLAs and their
clustering properties. A few years later, Lanzetta et al.
(1995) extended their work to include DLAs at lower redshift
by exploiting ultraviolet data obtained from the 
{\it International Ultraviolet Explorer} ({\it IUE})
satellite to gain one of the first glimpses into the
$z_{\rm abs} < 1.6$ absorber population.  These authors found that
$\Omega_{\rm DLA}$ decreases significantly from $z \sim 3.5$ to
$z \sim 0.01$, and interpreted this redshift evolution as 
evidence for the consumption of H~I gas by star formation. 

Wolfe et al. (1995) (WLFC95) used the Large Bright Quasar
Survey (LBQS, Hewett, Foltz \& Chaffee 1995 and references
therein) to search for DLAs with $1.6 \simlt z_{\rm abs}
\simlt 3.0$.  A total of 59 DLA candidates out of 228
spectra were pre-selected as having $W_0 \ge 5$ \AA.
At the time of publishing their paper,  Wolfe
et al. (1995) had confirmed the identification of 13 DLAs
with $N$(H~I)$ \ge 2 \times 10^{20}$ \cm\ from 15 candidates
with $W_0 > 9$ \AA.  In addition, there were 8 DLA
candidates whose  equivalent widths exceeded 10 \AA\ and
were therefore considered highly likely to be damped
systems.   All of these 8 candidates have subsequently been
confirmed as DLAs with intermediate resolution spectroscopy
(Storrie-Lombardi \& Wolfe 2000).  In addition to the LBQS
sample, WLFC95 constructed a `statistical sample' from the
literature consisting of 80 DLAs. From this they 
confirmed the coincidence which had previously been noted
(e.g. Lanzetta 1993) that the mass of H~I in DLAs at $z \sim
3$ is similar to the total luminous mass in stars today.
Expressing both as fractions of the closure density,
$\Omega_{\rm DLA} (z = 3.25) = (5.1 \pm 1.7) \times 10^{-3}
h^{-1}_{50}$ and $\Omega_{\star}(z = 0) \simeq 5 \times 
10^{-3} h^{-1}_{50}$ (e.g. Pagel 2000).

In order to extend studies of DLAs to lower redshifts, Rao
\& Turnshek (2000) have recently published results from an
observing campaign with the {\it Hubble Space Telescope}
({\it HST}).  Space-based telescopes are required in order to detect
low $z$ DLAs since the \lya\ signature at $z_{\rm abs} \simlt 1.6$
falls below the atmospheric cut-off.  Since such satellite resources
are limited, a pre-selection based on Mg~II and
Fe~II line strengths determined from ground-based
observations has permitted an efficient screening of the initial
QSO sample.  A total of 12 DLAs with $z_{\rm abs} < 1.5$
was found, thereby significantly improving the statistics
in this redshift regime. The puzzling result obtained from
these new data is that, contrary to previous indications,
$\Omega_{\rm DLA}$ appears to remain approximately constant
over all redshifts sampled, from $z = 3.5$ to 0.1\,. Using
{\it HST} archival spectra, Churchill (2001) has also argued
that the number density of DLAs implied from the incidence
of Mg~II systems remains unaltered down to $z_{\rm abs} \sim
0.05$. 

Two recent surveys have pushed the bounds to higher
redshifts. Storrie-Lombardi \& Wolfe (2000) and P\'{e}roux
et al. (2001a) have recently identified 19 and 15 DLAs
respectively at $z_{\rm abs} > 3.5$, including the highest
redshift damped absorber known to date ($z_{\rm abs} = 4.466$,
P\'{e}roux et al. 2001a; Dessauges-Zavadsky et al. 2001). 
Again, there is only
marginal evidence for evolution; between $z = 3.5$ and 4.5
$\Omega_{\rm DLA}$ is lower than at more recent epochs, but the
effect is only at the $\sim 2 \sigma$ level. 

The intention of this (non-exhaustive) review of previous DLA surveys is to
illustrate the difficulty of extracting unbiased statistics
from samples based on inhomogeneous datasets.  Many of these
surveys have built on existing samples, taken additional
targets from the literature and sometimes have insufficient
resolution to reliably determine whether an absorber is
damped. The advantage of this approach has been largely a
statistical one. The aggregate samples often contain several
hundred QSOs and such a large number is indeed required in
order to determine meaningful statistics because of the rarity of damped
systems.  The purpose of CORALS is to provide the first
complete and homogeneous survey for DLAs free from any bias
that may be inherent in magnitude-limited QSO samples.

\section{The CORALS Survey Sample}\label{def_sec}

We have based our new survey for DLAs on the radio-selected
Parkes quarter-Jansky sample (Shaver et al. 1996; Jackson et al. in prep.; 
Hook et al. in prep.).  This complete sample of flat-spectrum radio
sources was taken from the Parkes Catalogue (Wright \& Otrupcek 1990),
which consists of radio and optical data for almost 8300 radio sources 
for the sky south of declination +27deg.  Over most
of its area, it is complete to $S_{\rm 2.7}$ GHz = 0.25~Jy.
The Parkes quarter-Jansky sample
consists of all (878 in total) flat-spectrum ($\alpha > -0.4$, 
measured at 2.7 and 5.0 GHz) sources with
declinations between +2.5$^{\circ}$ and $-80^{\circ}$,
excluding low galactic latitudes ($\mid~b~\mid < 10^{\circ}$) and 
regions around the Magellanic Clouds.
The sample is complete to $S_{\rm 2.7}$ GHz = 0.25 Jy over 4.39 sr 
and to 0.60 Jy over a further 1.16 sr.  

Using this Parkes compilation of radio sources as the parent sample,
optical identification and $B$-band magnitudes for all targets 
with $\delta > -45^{\circ}$ were
achieved using the COSMOS Southern Sky Catalogue (Drinkwater, 
Barnes \& Ellison 1995) in the first instance or imaging at the 
ESO 3.6-m telescope at La Silla for the faintest QSOs.
Therefore, optical counterparts were identified for every radio
source in the Parkes compilation.
From this complete sample of 878 sources with no optical
magnitude limit or other selection bias, low resolution
spectra (FWHM = 12 -- 14 \AA) were obtained for the 442
stellar identifications (QSOs and BL Lacs) with the EFOSC on
the ESO 3.6-m to determine redshifts.  We note that since
compiling the CORALS sample, the catalogue of PKS sources has been
revised slightly to include some extra targets (Jackson et al. in prep.; 
Hook et al. in prep.).  However, it is important to stress that
this will not affect the results presented here, since the important
factor in our sample definition is that it is optically complete.

In our ground-based search for DLAs in this sample we
restricted ourselves to QSOs with $z_{\rm em} \geq 2.2$, so
as to be able to record a sufficient portion of the spectrum
blueward of the Ly$\alpha$ emission down to the onset of the
atmospheric cut-off. This left us with
a final sample of 66 QSOs in which we could search for DLAs
in the range $1.8 < z_{\rm abs} < z_{\rm em}$ with
sufficient sensitivity to measure reliable values of the
column density $N$(H~I).\footnote{One of the observed QSOs
actually has \ze\ $< 2.2$ due to an incorrect redshift in
our original list, see Table \ref{all_qsos}.}

\section{Observations and Spectral Analysis}\label{obs_sec}

Before commencing our observing campaign, we first searched
the literature to determine which QSOs had already been
observed at sufficiently high spectral resolution, S/N and
over the correct wavelength range to match the rest of our
DLA survey. The eight QSOs for which adequate spectra were
found in the literature were not re-observed. The remaining
58 QSOs were subject to an extensive observing campaign that
spanned five semesters on three different telescopes;
relevant details are presented in Table \ref{obs_log}.
References for the eight QSO spectra observed by others are
given in Table \ref{all_qsos}.  

\begin{table}
\begin{center}
\begin{tabular}{llcc} \hline \hline
 & & & \\
Telescope & Date & Resolution  & No. of \\
  &  (no. nights) & (at 4000 \AA) & QSOs observed \\
 & & & \\ \hline
 & & & \\
ESO 3.6-m & Sept. 1998 (2) & 6 -- 8 \AA & 16 \\
ESO 3.6-m & Feb. 1999 (2) & 7 \AA & 12 \\
AAT & Dec. 1998 (3$^a$) & 3 \AA & 11 \\
AAT & Apr. 1999 (2.5$^b$) & 3 \AA & 10 \\
AAT & Oct. 1999 (2) & 3 \AA & 12 \\
VLT & Oct. 2000 (0.5$^c$) & 4.5 \AA & 9 \\
VLT & Mar. 2001 (2) & 4.5 \AA & 10 \\
VLT & June 2001 (0.5$^d$) &  4.5 \AA & 1 \\
 & & & \\ \hline
\end{tabular}
\caption{\label{obs_log} Summary of observations CORALS survey.  
Many of the bright targets were
observed first with low resolution at the ESO 3.6-m and QSOs with
strong absorbers followed-up with AAT observations, whilst others
were observed directly with the AAT.  All faint targets ($B > 20$) 
were observed only with the VLT. 
Notes: 
$^a$ -- One night lost to bad weather. 
$^b$ -- The April AAT run consisted of 5 half nights.
$^c$ -- 6 hours of service time.
$^d$ -- 6 hours of Director's discretionary time.}
\end{center}
\end{table}

Our observing strategy has been to obtain `snapshot' spectra of
as many bright
(typically $B < 20$) QSOs as possible at low resolution with the
ESO 3.6-m in order to pre-select targets with candidate DLAs for
higher resolution follow-up at the AAT.
Targets fainter than this limit were observed directly with the VLT.
We briefly review the main characteristics of the observations.

\begin{table*}
\begin{center}
\begin{tabular}{lcccccc} \hline\hline
 & & & & & & \\
QSO & $z_{\rm em}$ & $B$ Mag & DLA? & $N$(H~I)
($10^{20}$\,cm$^{-2}$) & $z_{\rm abs}$ &
Ref \\
 & & & & & & \\ \hline
 & & & & & & \\
B0017$-$307 & 2.666 &  19.0 & N & ... & ... & \\   
B0039$-$407 & 2.478 &  18.5 & N & ... & ... & \\   
B0104$-$275 & 2.492 &  18.5 & N & ... & ... & \\   
B0113$-$283 & 2.555 &  19.0 & N & ... & ... & \\   
B0122$-$005 & 2.280 &  18.5 & N & ... & ... & \\  
B0244$-$128 & 2.201 &  18.5 & N & ... & ... & \\  
B0256$-$393 & 3.449 &  19.6 & N & ... & ... & \\  
B0325$-$222 & 2.220 &  19.0 & N & ... & ... & \\  
B0329$-$255 & 2.685 &  17.1 & N & ... & ... & 1\\  
B0335$-$122 & 3.442 &  21.5 & Y & 6.0 & 3.178 &  \\
B0347$-$211 & 2.944 &  21.1 & Y & 2.0 & 1.947 &  \\
B0405$-$331 & 2.570 &  19.0 & Y & 4.0 & 2.570$^a$ & \\  
B0420+022   & 2.277 &  19.5 & N & ... & ... &  \\
B0422$-$389 & 2.346 &  18.0 & N & ... & ... & \\  
B0432$-$440 & 2.649 &  19.6 & Y & 6.0 & 2.297 & \\  
B0434$-$188 & 2.702 &  20.0 & N & ... & ... & \\  
B0438$-$436 & 2.863 &  19.5 & Y & 6.0 & 2.347 &  \\ 
B0451$-$282 & 2.560 &  19.0 & N & ... & ... & \\  
B0458$-$020 & 2.286 &  20.0 & Y & 45.0& 2.039 & 2\\  
B0528$-$250 & 2.765 &  19.0 & Y & 5.6 & 2.141 & 3 \\  
            &       &       & Y & 15.8& 2.811$^{a}$ &4 \\  
B0537$-$286 & 3.110 &  20.0 & Y & 2.0 & 2.974 &  \\ 
B0601$-$172 & 2.711 &  20.0 & N & ... & ... & \\ 
B0610$-$436 & 3.461 &  19.0 & N & ... & ... &5 \\ 
B0819$-$032 & 2.352 &  18.2 & N & ... & ... & \\ 
B0834$-$201 & 2.752 &  19.0 & N & ... & ... & \\ 
B0913+003   & 3.074 &  21.7 & Y & 5.5 & 2.744 &  \\ 
B0919$-$260 & 2.300 &  19.0 & N & ... & ... & \\ 
B0933$-$333 & 2.906 &  20.0 & Y & 3.0 & 2.682 & \\ 
B1010$-$427 & 2.954 &  17.5 & N & ... & ... & \\ 
B1055$-$301 & 2.523 &  19.5 & Y & 35.0& 1.904& \\
B1136$-$156 & 2.625 &  20.0 & N & ... & ... & \\ 
B1147$-$192 & 2.489 &  19.4 & N & ... & ... &  \\ 
B1149$-$084 & 2.370 &  18.5 & N & ... & ... & \\ 
B1228$-$113 & 3.528 &  22.0 & Y & 4.0 & 2.193 &  \\ 
B1228$-$310 & 2.276 &  19.0 & N & ... & ... & \\ 
B1230$-$101 & 2.394 &  19.8 & Y & 3.0 & 1.931&  \\ 
B1251$-$407 & 4.464 &  23.7 & Y & 4.0 & 3.533 &  \\ 
            &       &       & Y & 2.0 & 3.752 &  \\ 
B1256$-$243 & 2.263 &  19.5 & N & ... & ... & \\ 
B1318$-$263$^b$ & 2.027 &  20.4 & N & ... & ... &  \\ 
B1351$-$018 & 3.710 &  20.9 & N & ... & ... &  \\ 
B1354$-$107 & 3.006 &  19.2 & Y & 2.5 & 2.501 & \\ 
            &       &       & Y & 6.0 & 2.966$^a$ & \\ 
B1402$-$012 & 2.518 &  18.2 & N & ... & ...&  \\ 
B1406$-$267 & 2.430 &  21.8 & N & ... & ...&   \\ 
B1418$-$064 & 3.689 &  18.5 & Y & 2.5 & 3.449 & \\ 
B1430$-$178 & 2.331 &  19.0 & N & ... & ...&  \\ 
B1535+004   & 3.497 &  24.1 & N & ... & ...&   \\ 
B1556$-$245 & 2.813 &  18.5 & N & ... & ...&  \\ 
B1635$-$035 & 2.871 &  21.8 & N & ... & ...&   \\ \hline
\end{tabular}
\caption{\label{all_qsos} Complete list of all CORALS QSOs and the
tally of DLAs identified.
References refer to QSOs for
which optical spectroscopy had already been obtained prior
to our survey, as follows: 1 -- Lanzetta
et al. (1991); 2 --  Pettini et al. (1997); 3 -- Ledoux et al. (1998);  
4 -- Lu et al. (1996); 5 -- M. Giavalisco, private communication;
6 -- Tytler, Fan and Burles (1996); 7 -- Songaila (1998);
8 -- Turnshek et al. (1989)
Notes: 
$^a$ -- $z_{\rm abs} \sim z_{\rm em}$; $^a$ -- The original redshift
of this target was incorrect in our list, the VLT spectrum confirms the
true redshift which is less than the survey cut-off at $z_{em} =2.2$. }
\end{center}
\end{table*}

\begin{table*}
\begin{center}
\begin{tabular}{lcccccc} \hline\hline
 & & & & & & \\
QSO & $z_{\rm em}$ & $B$ Mag & DLA? & $N$(H~I)
($10^{20}$\,cm$^{-2}$) & $z_{\rm abs}$ &
Ref \\
 & & & & & & \\ \hline
 & & & & & & \\
B1701+016   & 2.842 &  21.7 & N & ... & ...&   \\ 
B1705+018   & 2.575 &  18.9 & N & ... & ...&  \\  
B1937$-$101 & 3.780 &  19.0 & N & ... & ...&6  \\
B2000$-$330 & 3.780 &  18.5 & N & ... & ...&7  \\ 
B2126$-$158 & 3.275 &  17.5 & N & ... & ...&7  \\ 
B2149$-$307 & 2.330 &  17.5 & N & ... & ...&  \\ 
B2212$-$299 & 2.703 &  17.8 & N & ... & ...&  \\ 
B2215+020   & 3.550 &  21.5 & N & ... & ...&   \\ 
B2224+006   & 2.248 &  21.7 & N & ... & ...& \\ 
B2245$-$059 & 3.295 &  19.5 & N & ... & ...&  \\ 
B2245$-$328 & 2.268 &  16.5 & N & ... & ...&  \\ 
B2256+017   & 2.663 &  19.0 & N & ... & ...&8  \\ 
B2311$-$373 & 2.476 &  18.5 & Y & 3.0 & 2.182 &  \\ 
B2314$-$340 & 3.100 &  18.5 & N & ... & ... & \\ 
B2314$-$409 & 2.448 &  18.0 & Y & 4.0 & 1.857&  \\ 
            &       &       & Y & 2.0 & 1.875&  \\ 
B2315$-$172 & 2.462 &  19.5 & N & ... & ... &  \\ 
B2325$-$150 & 2.465 &  19.5 & N & ... & ... &  \\ 
B2351$-$154 & 2.665 &  19.0 & N & ... & ... & \\ 
 & & & & & & \\ \hline
\end{tabular}
\end{center}
\addtocounter{table}{-1}
\caption{Continued }
\end{table*}

\subsection{ESO 3.6-m Observations}

We obtained low resolution spectra of 28 QSOs using 
the EFOSC2 spectrograph on the ESO 3.6-m telescope during four nights
between Sept. 1998 and Feb. 1999. The spectral resolution
is approximately 7 \AA\ with a 1 arcsec slit and B600 grism, covering 
3270 -- 5240 \AA.
For a few QSOs, we also took spectra using R600 grism with the similar
spectral resolution to cover 4320 -- 6360 \AA. The seeing conditions
remained around 1.2 arcsec, although the sky
was not photometric during all four nights. However, since we normalise
our spectra to fit the DLA profile, the accurate flux calibration is not
critical for the purpose of this paper. 

\subsection{AAT Observations}

The RGO spectrograph was used with the TEK CCD, 25cm camera
and 600~V grating.  The seeing at Siding Spring was highly
variable through these observations, ranging from 0.8 to 2.0
arcsecs, often with large fluctuations during a given night.
However, the slit width was fixed at 1.5 arcsec which
projected onto 1.8 pixels to give a FWHM resolution of
3\,\AA.  The grating angle was chosen to cover the entire
wavelength range of each QSO from at least 3400 \AA\
(corresponding to the wavelength of \lya\ $\lambda 1216$ at
$z_{\rm abs} = 1.8$) to $z_{\rm em}$.  The grating angle
varied slightly for each run depending on the QSO subset to
be observed, but was typically around $19.74^{\circ}$,
corresponding to a central wavelength of $\lambda =
4000$\,\AA.  At this setting, the wavelength range is $\sim
3200$\,\AA\ -- 4800\,\AA.  For the few bright, high redshift
($z_{\rm em} \simgt 2.9$) targets observed with the AAT, a
second grating setting, with a central wavelength $\lambda =
5500$\,\AA,  was required to cover the spectrum up to 6300\,
\AA.

\subsection{VLT Observations}

The VLT observations were executed with FORS1 in a
combination of service and visitor mode.  The 600B grating
was used for all targets, with additional wavelength
coverage provided by the 600R grating for the $z_{\rm em} =
4.458$ QSO B1251$-$407.  A slit width of 0.7 arcsec provided
a FWHM resolution of 4.5 \AA\ and a wavelength coverage of
3360 -- 5760\,\AA.  The 600R grating gave additional
coverage (only required for the highest redshift QSO in our
sample) over 5200 -- 7300\,\AA\ with a resolution of 3.6 \AA.
Despite the faintness of several of the VLT targets (down to
$B = 24$), all acquisitions could be executed in `fast' mode
and without blind offsets.

\subsection{Data Reduction and Column Density Fitting}

We applied the same reduction procedure to all of the data.
The standard IRAF\footnote{IRAF is distributed by the
National Optical Astronomy Observatories, which are operated
by the Association of Universities for Research 
in Astronomy, Inc, under cooperative agreement with the 
National Science Foundation.} routines were used.
First, all of the images were trimmed and 
the bias level was subtracted using the over-scan regions.
High S/N flat-field images were combined to a single
image, which was then smoothed using a box median filter of 
1 $\times$ 50 pixels. The orientation of the box median filter is 
chosen such that we preserve all of the variations along the dispersion axis.
Then, the normalized 2-D flat-field image was used to remove pixel-to-pixel
variations in the quasar images. The task APALL was used to perform the 
optimal extraction of the 1-D spectra. 
Wavelength calibration images were typically taken both before and after 
each QSO image. The comparison lamps
are CuAr at AAT, NeHe at ESO 3.6m and NeHgCd on FORS1 at VLT.
The dispersion solutions were obtained using a 4th order
Legendre polynomial and the RMS errors of the fitting was less than
0.05\AA. Finally, we performed error-weighted summation of 
all of the wavelength calibrated spectra for
each QSO. The final error array was the quadratic sum of the
individual error spectra.

The spectra of all 58 QSOs observed by us are presented in
Figure \ref{all_spec}. Once extracted, the spectra were
inspected for the presence of DLAs -- the \lya\ signature
clearly visible as a broad, saturated absorption feature.
If a DLA was identified, the spectrum was normalised by
dividing through by the QSO continuum and then fitted with a
\lya\ profile using the Starlink package DIPSO to determine
the redshift and column density of the DLA.  The
complete list of CORALS QSOs and identified DLAs can be
found in Table \ref{all_qsos}.

\begin{figure*}[h]
\centerline{\resizebox{16cm}{!}
{\includegraphics{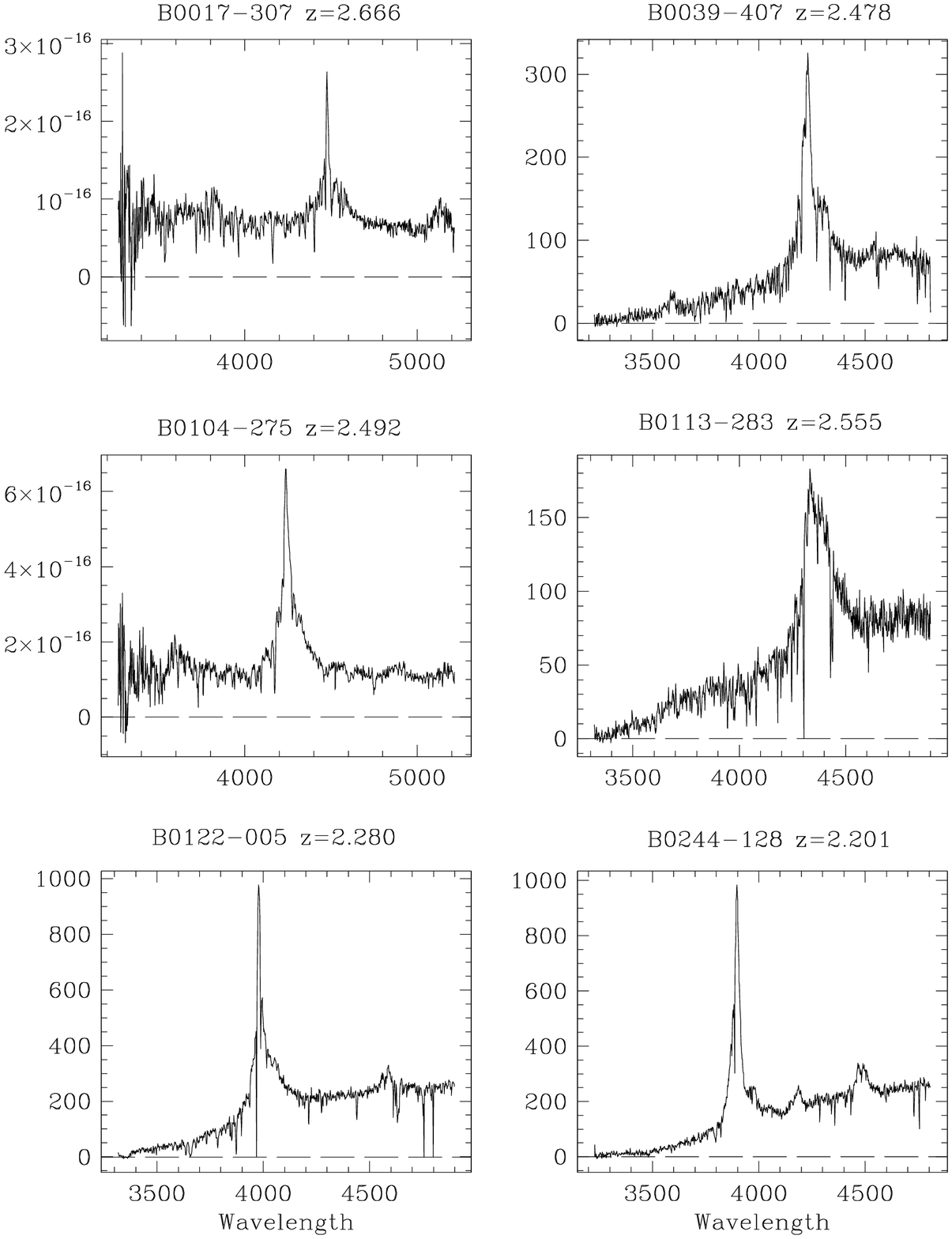}}}
\caption{\label{all_spec} Spectra obtained for the CORALS
QSOs at the AAT (RGO spectrograph), ESO 3.6-m (EFOSC), and
VLT (FORS1) facilities.  Note that neither the higher
resolution AAT spectra (FWHM = 3\, \AA) nor the VLT ones are
flux calibrated and the $y$-axis is simply relative counts
uncorrected for instrument response and atmospheric
transmission. The ESO 3.6-m data (FWHM $\sim$ 7\,\AA),
however, were flux calibrated and the $y$-axis scale is in
units of ergs s$^{-1}$ \cm\ \AA$^{-1}$.  Vertical tick marks show
the position and redshift of intervening DLAs, if present. }
\end{figure*}

\begin{figure*}[h]
\centerline{\resizebox{16cm}{!}
{\includegraphics{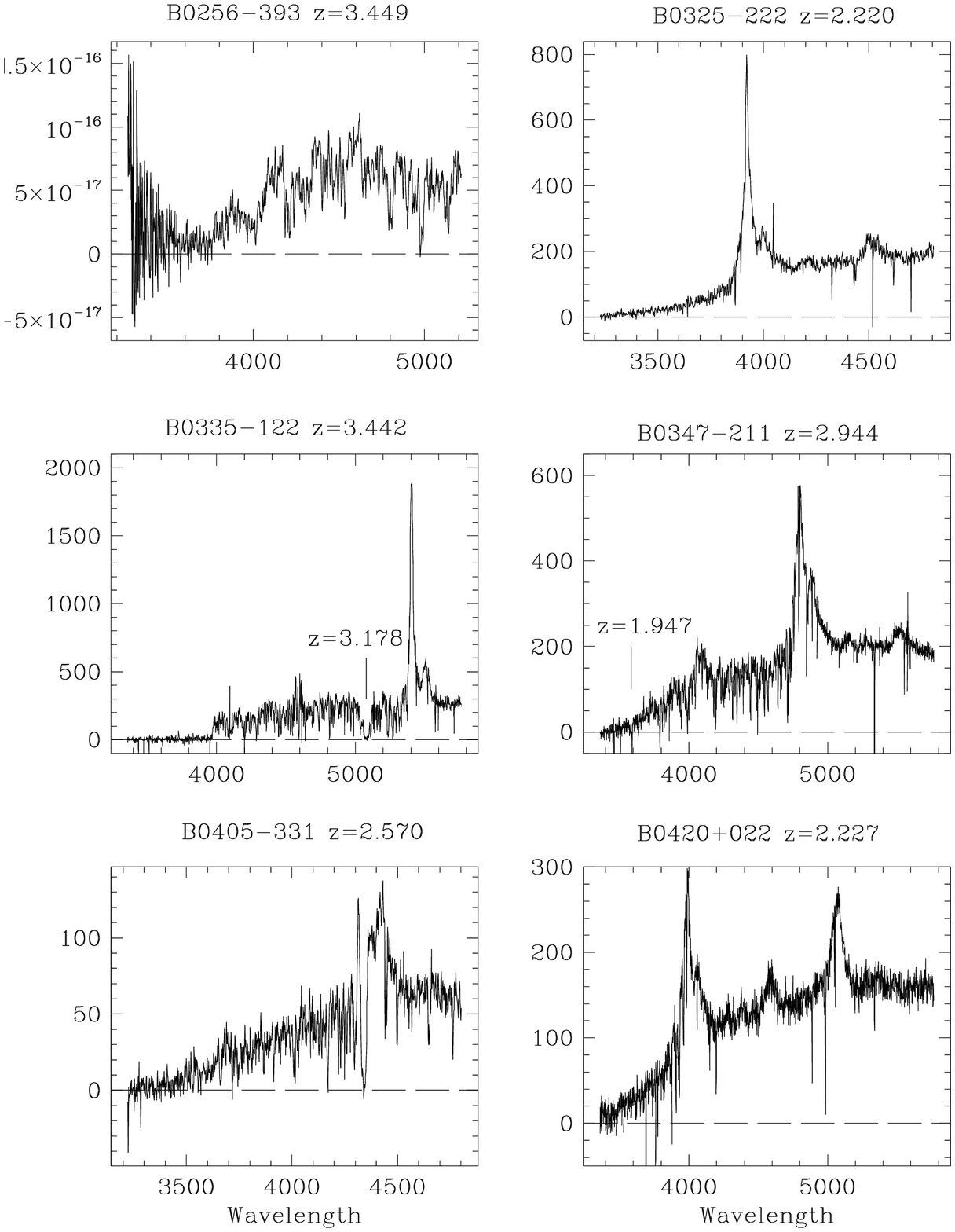}}}
\addtocounter{figure}{-1}
\caption{Continued }
\end{figure*}

\begin{figure*}[h]
\centerline{\resizebox{16cm}{!}
{\includegraphics{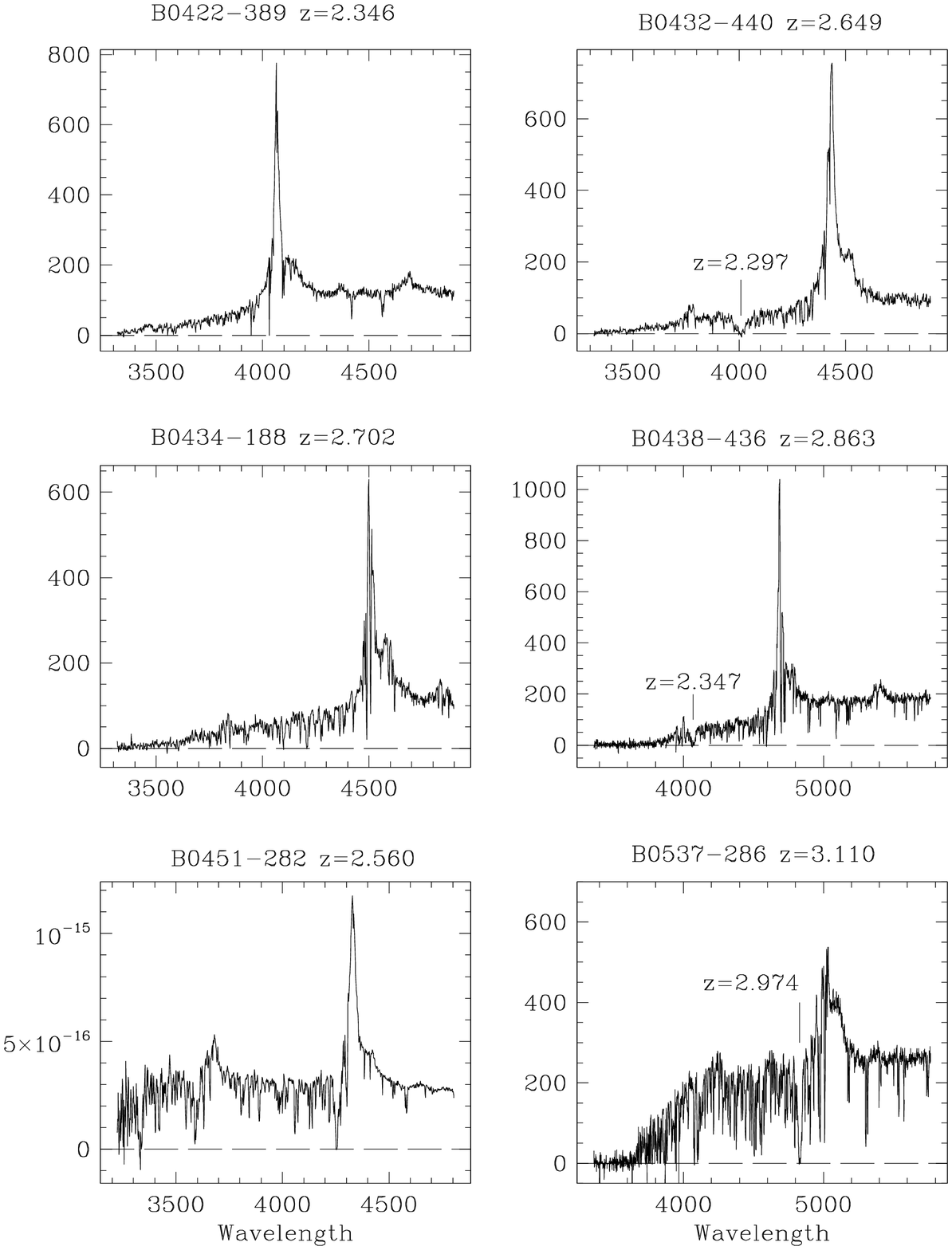}}}
\addtocounter{figure}{-1}
\caption{Continued }
\end{figure*}

\begin{figure*}[h]
\centerline{\resizebox{16cm}{!}
{\includegraphics{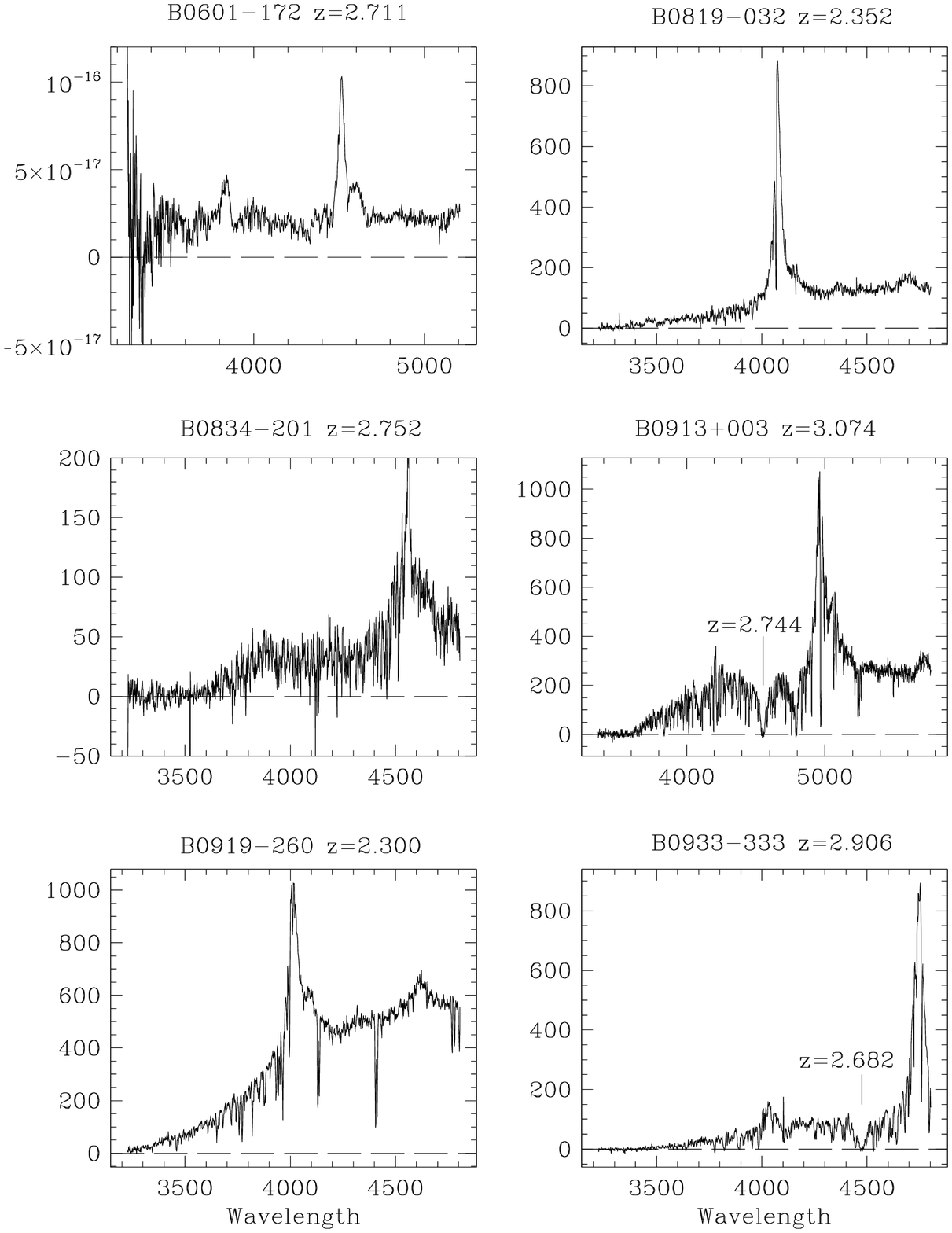}}}
\addtocounter{figure}{-1}
\caption{Continued }
\end{figure*}

\begin{figure*}[h]
\centerline{\resizebox{16cm}{!}
{\includegraphics{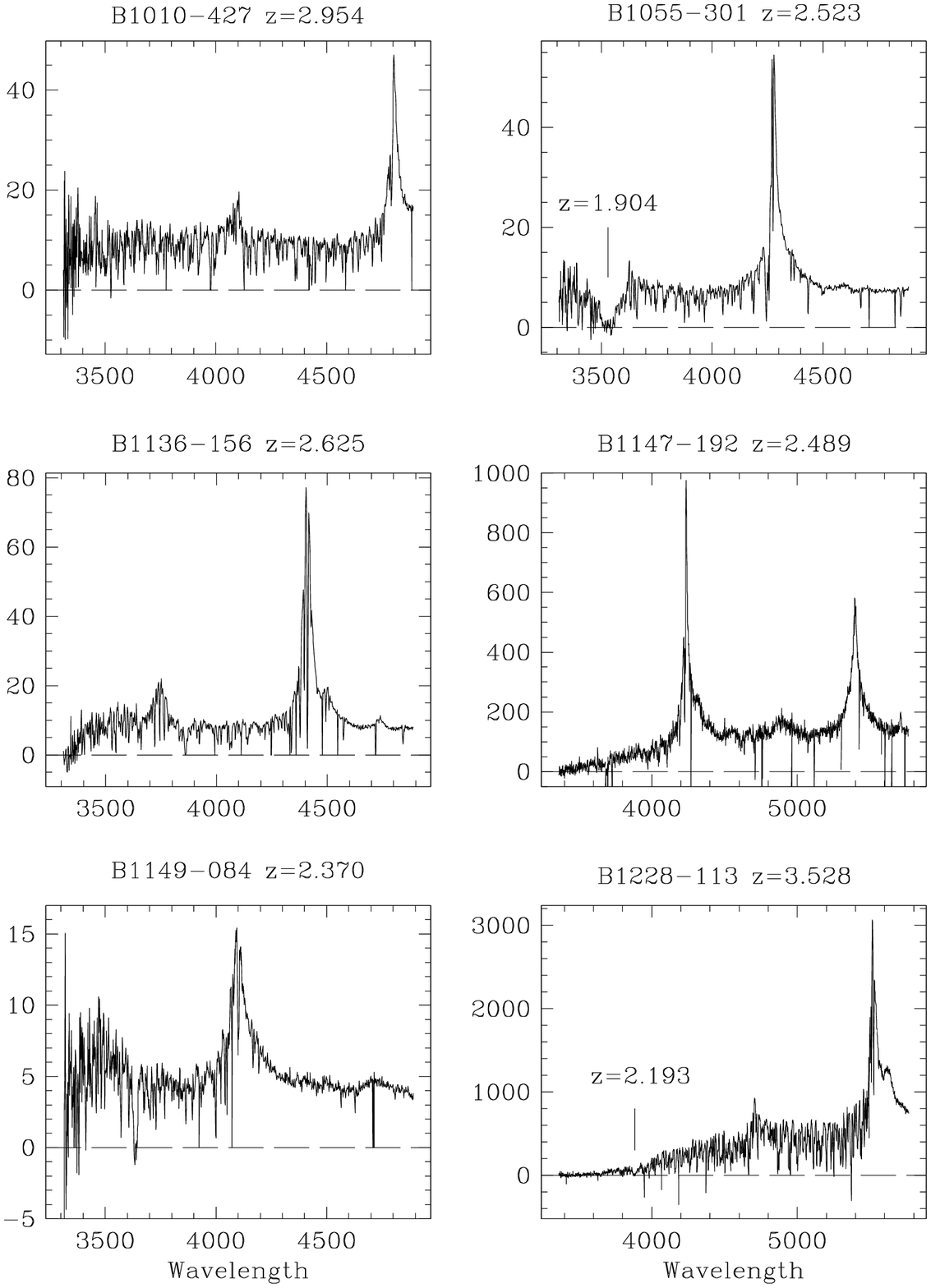}}}
\addtocounter{figure}{-1}
\caption{Continued }
\end{figure*}

\begin{figure*}[h]
\centerline{\resizebox{16cm}{!}
{\includegraphics{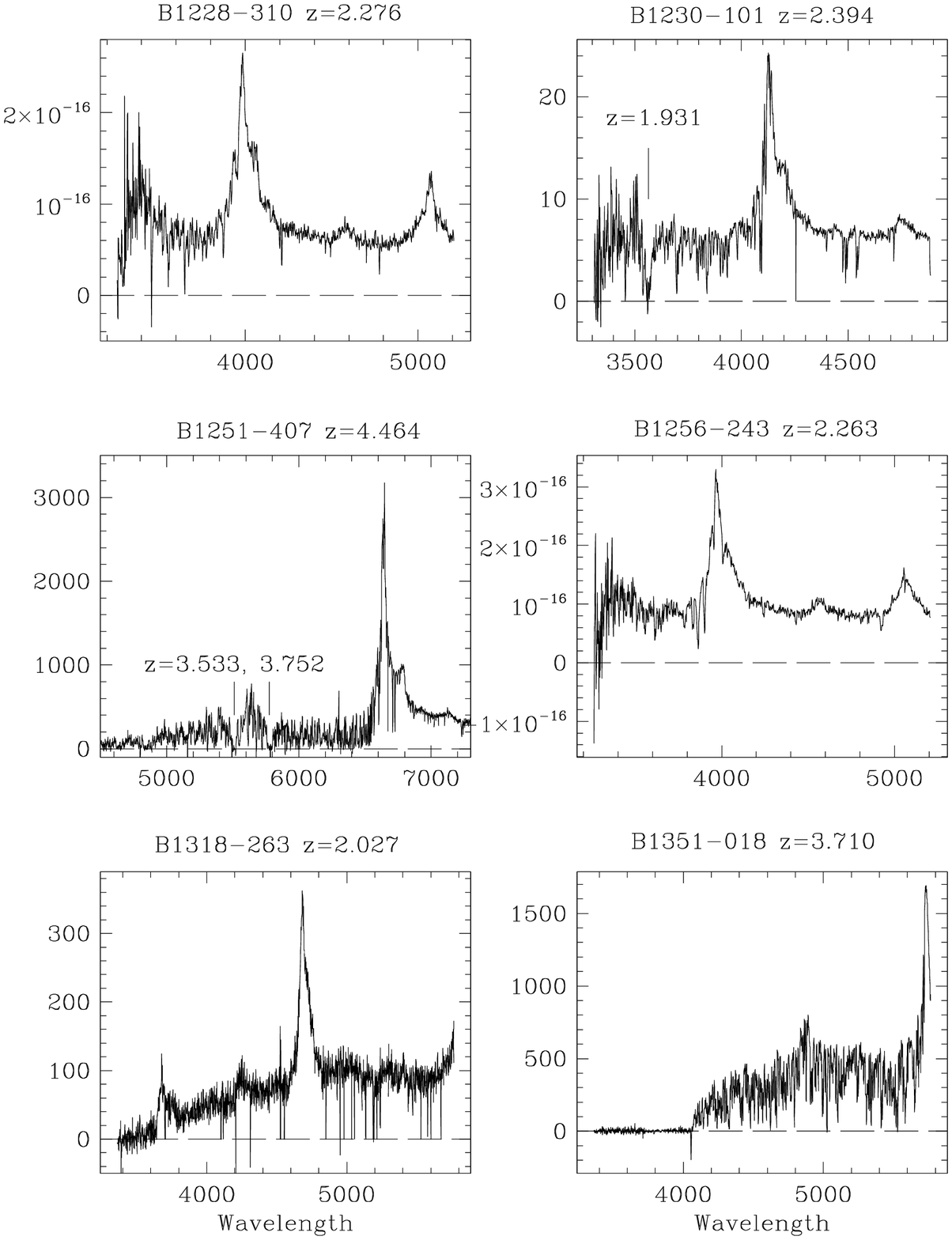}}}
\addtocounter{figure}{-1}
\caption{Continued }
\end{figure*}

\begin{figure*}[h]
\centerline{\resizebox{16cm}{!}
{\includegraphics{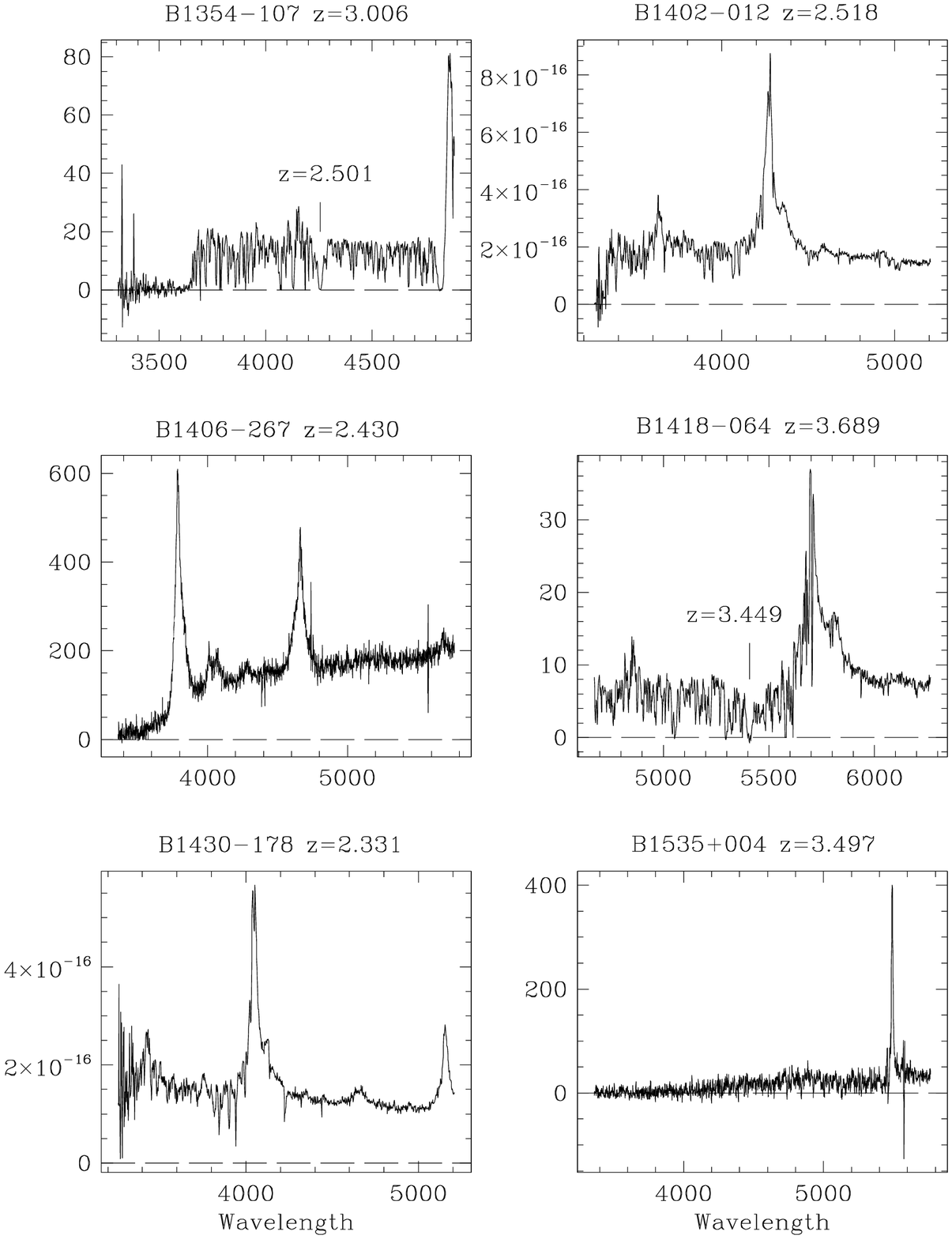}}}
\addtocounter{figure}{-1}
\caption{Continued }
\end{figure*}

\begin{figure*}[h]
\centerline{\resizebox{16cm}{!}
{\includegraphics{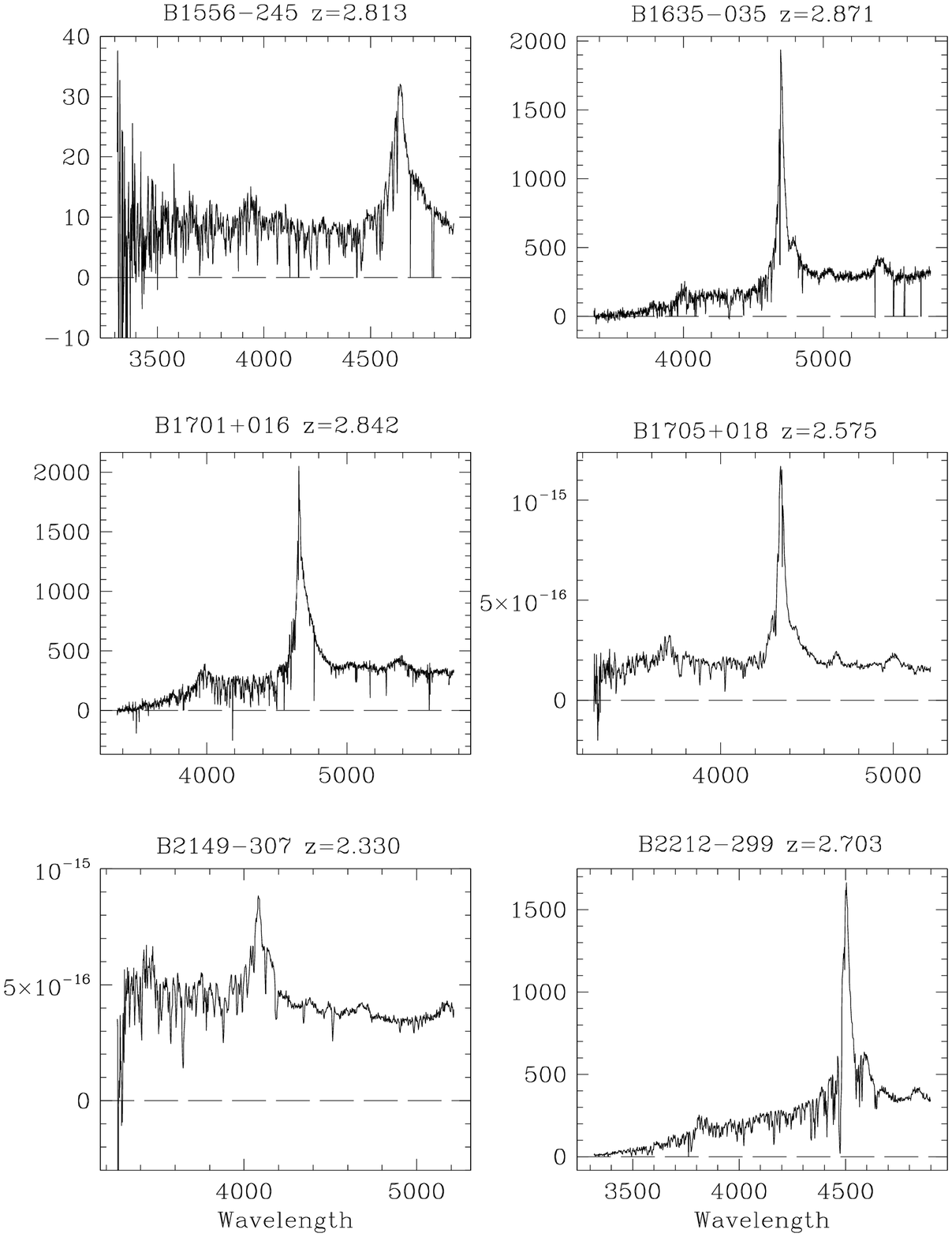}}}
\addtocounter{figure}{-1}
\caption{Continued }
\end{figure*}

\begin{figure*}[h]
\centerline{\resizebox{16cm}{!}
{\includegraphics{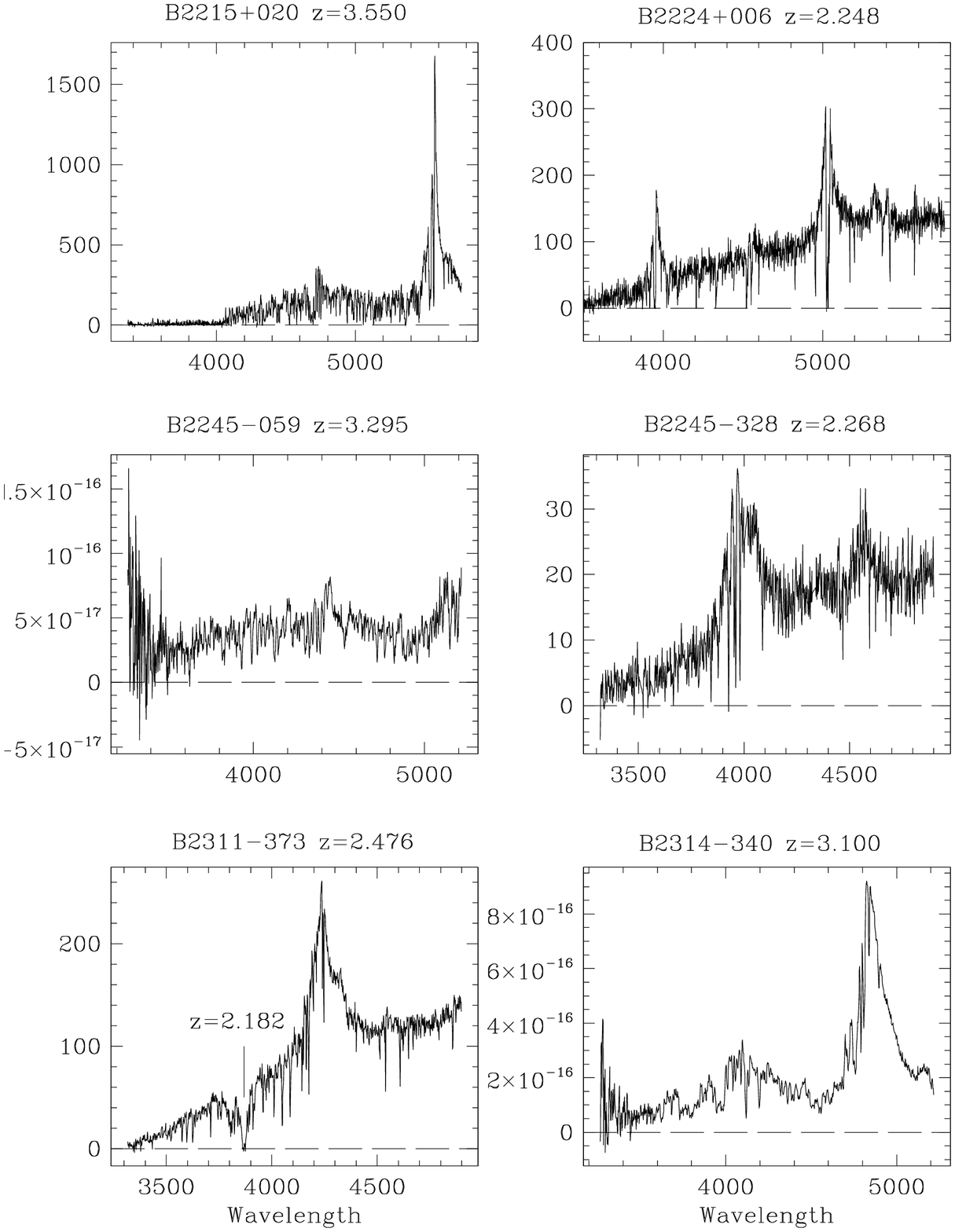}}}
\addtocounter{figure}{-1}
\caption{Continued }
\end{figure*}

\begin{figure*}[h]
\centerline{\resizebox{16cm}{!}
{\includegraphics{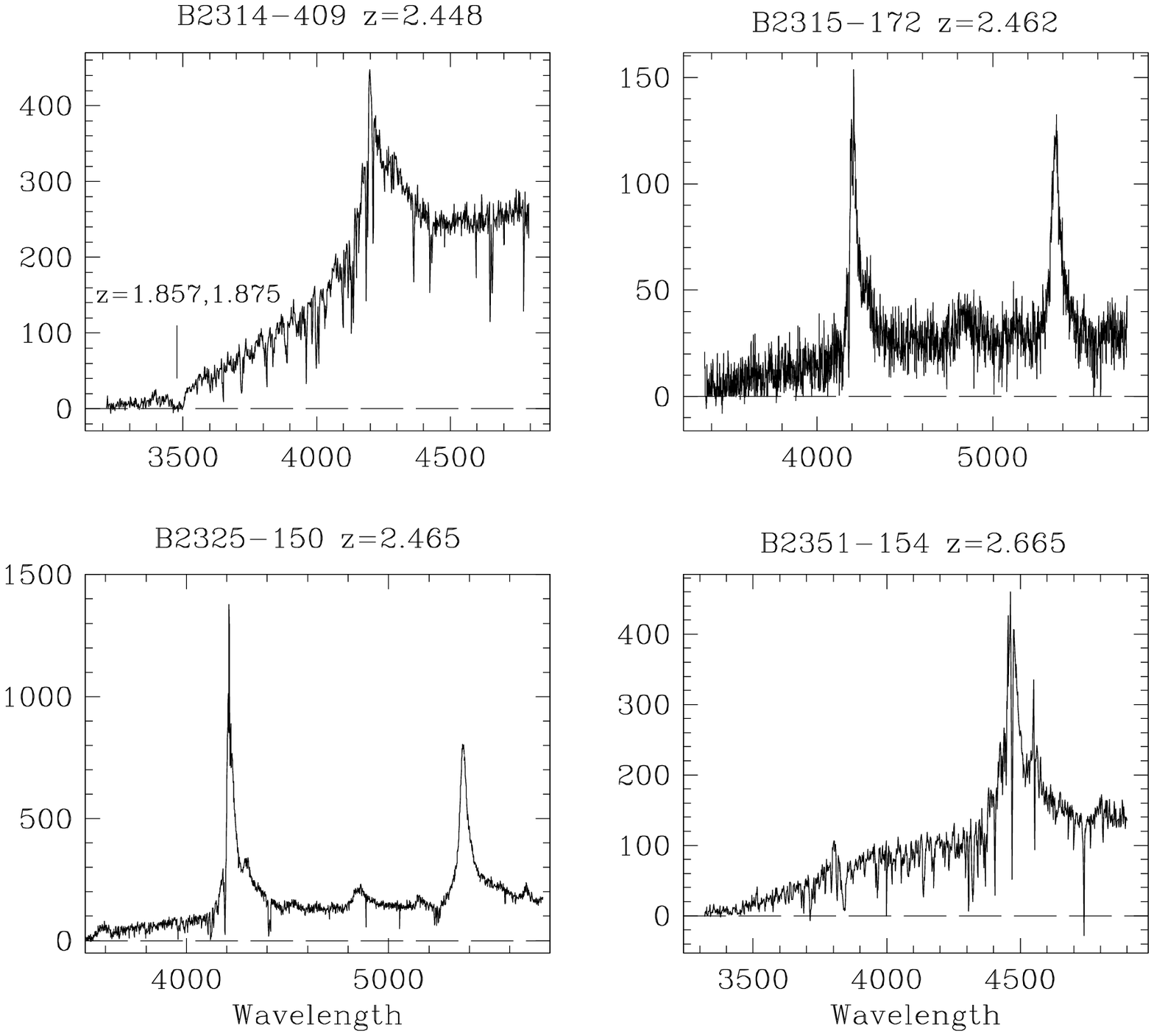}}}
\addtocounter{figure}{-1}
\caption{Continued }
\end{figure*}

\section{Damped Lyman Alpha Systems in the CORALS Survey}\label{dla_sec}

We adopt the usual definition of a DLA, i.e. \nhi\ $\ge 2
\times 10^{20}$ \cm, although our data are of sufficient
resolution and S/N to recognise and measure absorbers of
somewhat lower \nhi\ (Ellison 2000). From the observations
of the 66 CORALS QSOs, a total of 22 DLAs have been
identified. Three of these have absorption redshifts similar
to the emission redshift of the QSO. We follow the standard
procedure of excluding DLAs within 3000 \kms\ of the QSO
redshift from our statistical analysis, in order to
facilitate comparison with other surveys.  However, we note
that the $z_{\rm abs} \sim z_{\rm em}$ DLAs are probably similar to
intervening absorbers (M\o ller, Warren \& Fynbo 1998) and
we defer the analysis of the `associated' CORALS DLAs to a
future paper (Ellison et al., in preparation).  Of the 19
intervening DLAs, two (B0458$-$020 and B0528$-$250a) were
already known and have been extensively studied prior to our
survey. Profile fits to all the others are shown in Figure
\ref{all_DLAs}. We now briefly discuss each DLA system.

\begin{figure*}
\centerline{\resizebox{14cm}{!}
{\includegraphics{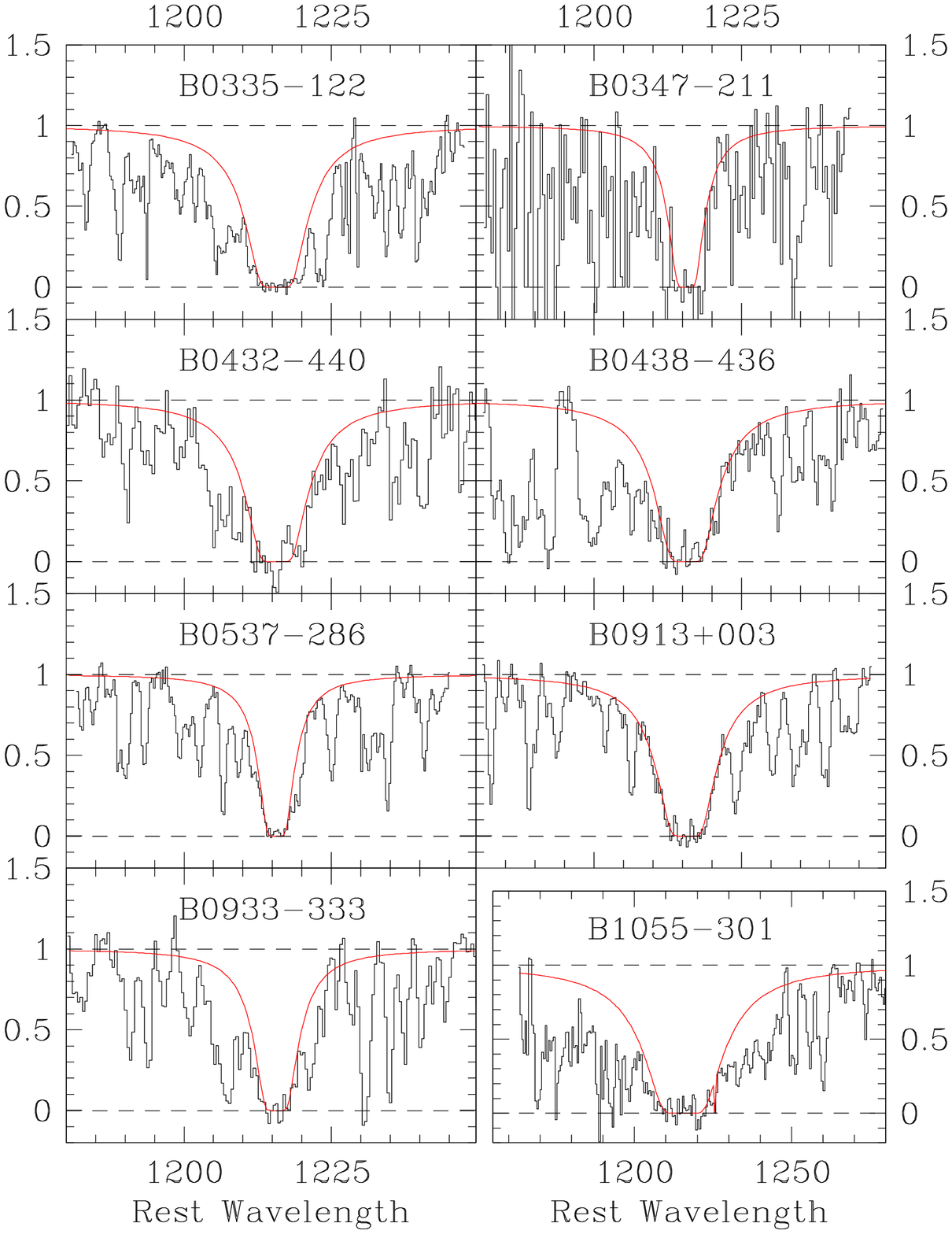}}}
\caption{\label{all_DLAs} Damped \lya\ profile fits (continuous lines) 
to all newly discovered 
intervening DLAs. See Table 3 for the values of \nhi\ and
$z_{\rm abs}$ corresponding to the theoretical profiles
shown.  Not that the bottom right-hand panel is on a different
wavelength scale due to the large column density of the DLA towards
B1055$-$301.
}
\end{figure*}

\begin{figure*}
\centerline{\resizebox{14cm}{!}
{\includegraphics{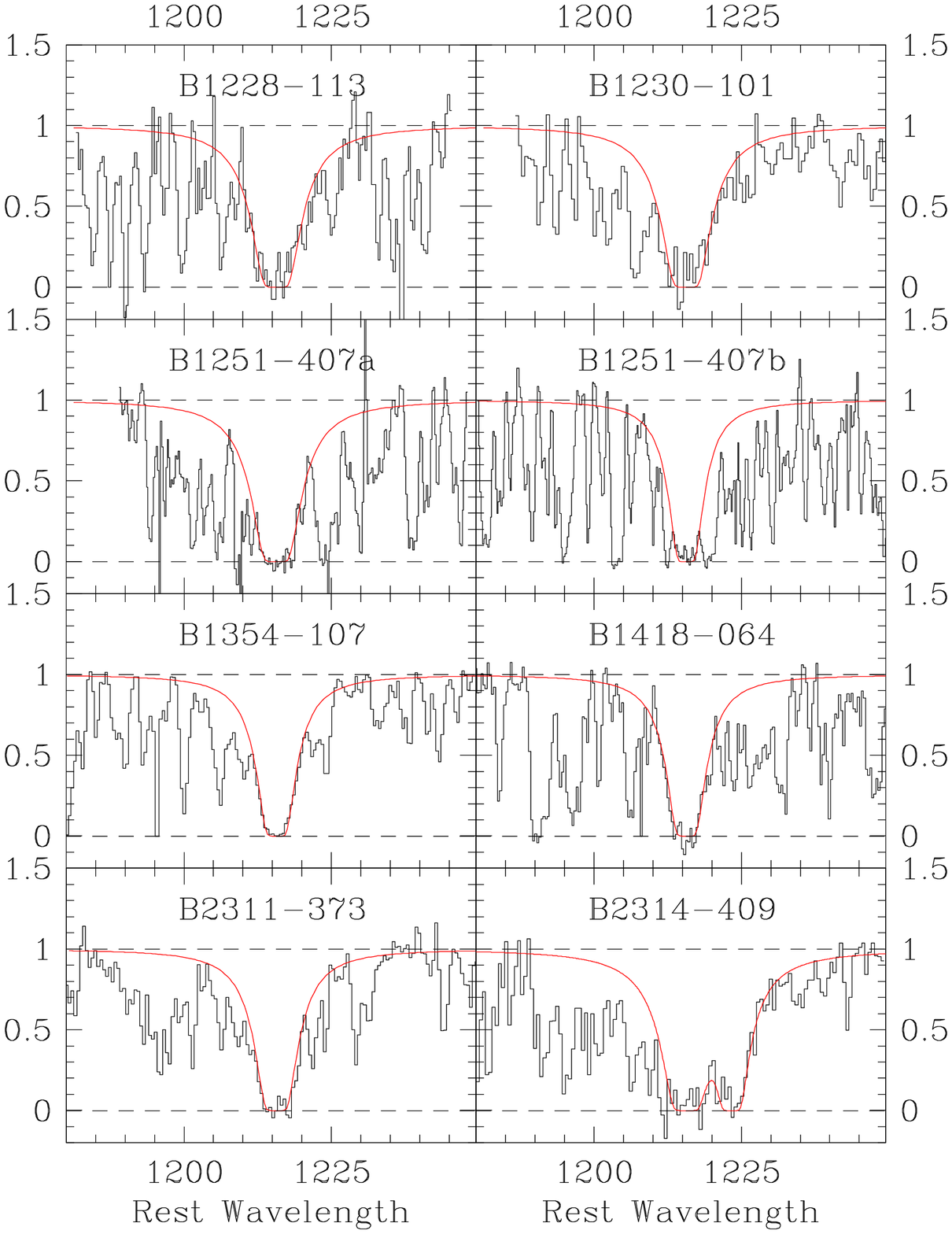}}}
\addtocounter{figure}{-1}
\caption{Continued }
\end{figure*}

\paragraph{B0335$-$122}  Extended absorption around the DLA trough complicates
the fitting of this system. The O~I $\lambda$1302 line is used to constrain 
redshift, but the error on the final fit is relatively high ($6\pm1 \times 
10^{20}$ \cm) due the critical, yet uncertain, continuum placement.  

\paragraph{B0347$-$211}  The combination of low absorption redshift and
faint QSO magnitude have resulted in a low S/N spectrum of this DLA and
a relatively poor fit.  The redshift is constrained by the Al~II $\lambda$1670 
line and the \nhi\ determined by fitting a damped profile is in good
agreement with the column density inferred from the equivalent width 
measurement.

\paragraph{B0432$-$440}  Although the spectrum is noisy, the column density of
this DLA is reasonably constrained by the base of its trough and the shape 
of its red wing.  This system is well-fitted with a column density $N$(H~I)
= $6 \times 10^{20}$ \cm.

\paragraph{B0438$-$436}  Both AAT and VLT spectra were obtained for
this QSO.   The best fit to the combined spectrum has \nhi = $6 \times 
10^{20}$ \cm, although this is constrained mostly by the red wing since
there is significant contaminating absorption in the blue wing.
The redshift of the Al~II $\lambda 1670$ line 
(with rest-frame equivalent width $W_0$ = 0.63\,\AA) agrees well
with the redshift determined from the centre of the DLA trough.

\paragraph{B0537$-$286} A relatively high S/N spectrum and simple structure
around the absorber permits a good fit to this DLA.

\paragraph{B0913$+$003}  An excellent fit to this DLA is facilitated by
the clearly defined damping wings and lack of blending.

\paragraph{B0933$-$333}  Despite blending with a weaker component
blueward of the DLA, this system is reasonably fit with 
profile of $N$(H~I) = $3 \times 10^{20}$ \cm.

\paragraph{B1055$-$301}  This DLA has a very
large equivalent width ($W_0 \sim 60$ \AA) but is heavily blended with
other absorption lines.  The fit is only constrained by the base of
the absorption which is clearly saturated over 10 \AA\ in the rest
frame.  Higher order Lyman lines are not available to improve the
decomposition of the H~I cloud model, although several metal
transitions are covered by the AAT spectrum. Strong Si~II
$\lambda$1526, Al~II $\lambda$1671 and Fe~II $\lambda$1608 are all
observed with redshifts of 1.9037 and rest frame equivalent widths
$W_0 = 1.2$\,\AA, 1.3 \AA\ and 0.8 \AA\ respectively.  This provides strong
support for the presence of 
a DLA at the position shown in Figure \ref{all_DLAs}.  However, the
column density can only be constrained to within 
$\pm 15$\%: \nhi = $35\pm5 \times 10^{20}$ \cm.

\paragraph{B1228$-$113} Another low redshift system whose spectrum
has a low S/N, although the lack of strong nearby \lya\ forest 
lines results in an acceptable fit.

\paragraph{B1230$-$101}  Constrained mostly by its fit to
the red wing, this DLA at $z_{\rm abs} = 1.931$ has several
metal lines associated with it.  The AAT spectrum covers
both Fe~II $\lambda$1608 ($W_0$ = 520 m\AA) and Si~II
$\lambda$1526 ($W_0$ = 720 m\AA); the latter lies just
blueward of a strong, resolved, C~IV doublet at $z_{\rm abs}
= 1.899$.  There is a second C~IV system at $\lambda_{\rm
obs} \sim 4540$ \AA\ associated with the DLA itself.

\paragraph{B1251$-$407a,b}  As can be seen in Figure
\ref{all_spec}, this QSO has two prominent absorption lines
at approximately 5510 and 5780\,\AA, corresponding to \lya\
at $z_{\rm abs} = 3.533$ and 3.752 respectively. The former
is well reproduced by a damped profile with \nhi\ = $4
\times 10^{20}$\cm. The latter has a large equivalent width
($W_0 = 13.7$\,\AA), but also steep sides and cannot be
fitted satisfactorily with a single absorption component.
Closer inspection reveals structure within the core of the
damped \lya\ line (see Figure 2). We consider this feature
to be a composite consisting of a DLA at $z_{\rm abs} =
3.752$ with \nhi\ = $2 \times 10^{20}$\cm\ flanked by two
lower column density components. This interpretation is 
supported by the presence of Si~II $\lambda$1526 absorption
at the same redshift as the DLA.  In any case, both DLAs in
this QSO are excluded from our discussion of the sample
statistics below because we consider only the redshift
interval $1.8 < z_{\rm abs} < 3.5$ (see \S\ref{stats_sec}).  The
two other moderately large EW systems at $\lambda = 4640$
and 4830\,\AA\ respectively (see Figure 1), are not DLAs,
but blends of lower column density lines.

\paragraph{B1354$-$107a}  Two strong absorption features are
seen towards this QSO.  Absorber `a' has a redshift of
$z_{\rm abs} = 2.501$ and is well fitted by a damped profile
with $N$(H~I)$ = 2.5 \times 10^{20}$ \cm.  Absorber `b',
with $N$(H~I)$ = 6 \times 10^{20}$ \cm and $z_{\rm abs} =
2.966$, is classified as a $z_{\rm abs} \sim z_{\rm em}$ DLA
and is not included in the present analysis.

\paragraph{B1418$-$064}  At $z_{\rm abs} = 3.449$, this is
the highest redshift DLA to be included in our statistical
analysis; the \lya\ line is well reproduced by a damped
profile with $N$(H~I) = 2.5$ \times 10^{20}$ \cm. Our
limited spectral coverage redward of \lya\ emission
reveals associated Si~II $\lambda$1304 and O~I $\lambda$1302
absorption lines with $W_0 = 290$ and 390\,m\AA\ respectively
(although the O~I transition is probably blended with an
unidentified line and the equivalent width measurement is
therefore an overestimate).  


\paragraph{B2311$-$373}  A column density of $N$(H~I) = 3 $\times
10^{20}$ \cm\ provides a satisfactory fit to the base and both
wings of this DLA.   Si~II $\lambda$1526 is the only metal transition
identified in the AAT spectrum with $W_0 = 500$\, m\AA. 

\paragraph{B2314$-$409a,b} As is the case for B1055$-$301,
there is extended H~I absorption in this spectrum. The spike
at $\lambda_0 \sim 1220$ \AA\ in Figure 2 could be either
noise or residual continuum flux between two closely spaced
components.  The latter interpretation is supported by the
presence of two Si~II $\lambda$1526 absorption lines at
redshifts $z_{\rm abs} = 1.857$ and 1.875 respectively.
These redshifts match well the blue and red components of
the \lya\ absorption feature, as indicated by the fit shown
in Figure 2. The metal lines are stronger in the $z_{\rm
abs} = 1.857$ system, where we also detect Fe~II~$\lambda
1608$ and Al~II~$\lambda 1671$; the $z_{\rm abs} = 1.875$
component may well be a very low metallicity system given
that Fe~II~$\lambda 1608$ has  $W_0 \leq 90$\,m\AA. This
is the second case of a multiple DLA, the first being a
triple DLA (CTQ247) spread over $\sim 6000$ \kms, discovered
by Lopez et al. (2000).

\section{DLA Statistics - Number Density and $\Omega_{\rm DLA}$}\label{stats_sec}

One of the key areas of interest in the study of DLAs is how
the population evolves with time.  If DLAs represent the
bulk of the galaxy population at a given epoch, then
studying their properties as a function of redshift will be
a powerful method for tracing galaxy evolution. For example,
there is strong evidence for redshift evolution in the 
column density distribution function, $f(N)$, 
(Storrie-Lombardi \& Wolfe 2000).  However, since the number
density is well represented by a power law there appears to
be no change in the product of space density and absorber
cross section. As discussed in the Introduction, the
evolution of $\Omega_{\rm DLA}$ is still unclear,
particularly at $z_{\rm abs} < 1.5$ where statistics are poor.

Due to the relatively small size of the CORALS survey, it is
not possible to investigate the evolution of DLA statistics
for this sample.  However, since our main objective is to
ascertain whether or not a significant fraction of gas has
gone undetected, it is sufficient for us to
restrict our determination of $\Omega_{\rm DLA}$ to the
range $1.8 < z_{\rm abs} < 3.5$ where there appears to be
little evolution.  This requires us to omit two DLAs from
our sample (B1251$-$407a and b), and restrict our
statistical analysis to the remaining 17 DLAs.

\subsection{DLA Number Density}

This is simply the total number of DLAs divided by the total
intervening redshift interval covered, $\Delta z$, given by 

\begin{equation} 
\Delta z = \sum_{i = 1}^{n} (z_{i}^{max} - z_{i}^{min}) 
\end{equation}

where the summation is over the $n$ QSOs in the sample. For
CORALS, $z_{\rm min} = 1.8$\footnote{We note that although most surveys
calculate $z_{\rm min}$ for each QSO on a case by case basis, depending
on wavelength coverage and S/N, we have designed the CORALS survey to
have adequate spectral range and S/N to achieve a uniform lower
redshift cut-off.} and $z_{max} = 3.5$ or $z_{3000}$ (the redshift
corresponding to a $v$ = 3000 \kms), whichever is smaller:

\begin{equation}
z_{3000} = \left[ (z_{em} + 1) \sqrt{ \frac{(~b - 1)}{(-b - 1)} } \right] - 1
\end{equation}

\noindent and

\begin{equation}
b = \frac{v}{c} = 0.01,
\end{equation}

In cases where a Lyman limit system is present in the QSO
spectrum within the redshift range considered,

\begin{equation}
z_{min} = \frac{(z_{LLS} + 1) \times 912}{1216} - 1
\end{equation}

The total redshift interval for the CORALS sample is $\Delta
z = 55.46$ within which we detect 17 DLAs. Thus, $n(z) =
0.31^{+0.09}_{-0.08}$ at a mean absorption redshift $\langle
z_{abs} \rangle= 2.37$  (the errors were calculated using
the prescription by Gehrels 1986 for small number
statistics). 

Taken at face value, the estimate of $n(z)$ from the CORALS
survey is $\sim 50$\% larger (at the
same  $\langle z_{abs} \rangle$) than that found in previous
surveys. For example, Storrie-Lombardi \& Wolfe (2000)
deduced $n(z)=0.055(1+z)^{1.11} = 0.21$ (no errors quoted).
However, this difference is only marginally
significant, since the two determinations of $n(z)$
are within $\sim 1 \sigma$ of each other.

\subsection{The Mass Density of Neutral Gas in DLAs}

\begin{figure}
\centerline{\resizebox{7.5cm}{!}
{\includegraphics{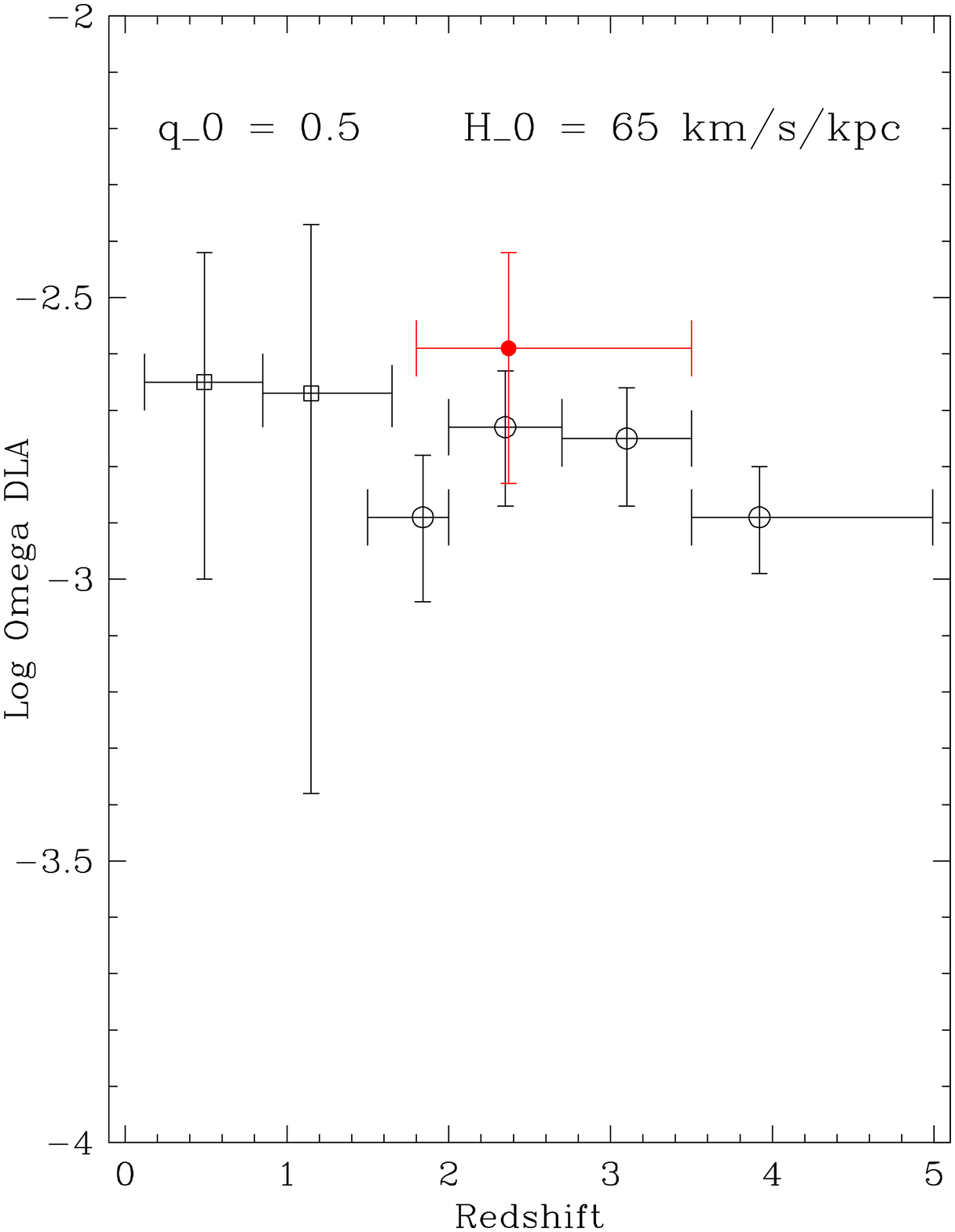}}}
\caption{\label{omega} The mass density of neutral gas, $\Omega_{\rm
DLA}$, in DLAs.  Open circles and squares are measurements
from the latest compilations by P\'{e}roux et al. (2001b) and
Rao \& Turnshek (2000) respectively.  The solid circle is
the value from the CORALS survey presented here for the
redshift interval $1.8 < z_{abs} < 3.5$. }
\end{figure}

The mass density of neutral gas in DLAs as a fraction of the closure
density is expressed as

\begin{equation}
\Omega_{\rm DLA} = \frac{H_0 \mu m_H}{c \rho_{crit}}
  \int_{N_{min}}^{N_{max}} N f(N) dN
\end{equation}

where $\mu$ is the mean molecular weight (= 1.3), $m_H$ is
the mass of a hydrogen atom, and $f(N)$ is the column
density distribution. We avoid making \textit{a priori}
assumptions about the functional shape of $f(N)$ by adopting
the approximation by Storrie-Lombardi, McMahon \& Irwin
(1996)

\begin{equation}\label{omega_approx}
\int_{N_{min}}^{N_{max}} N f(N,z) dN =\frac{ \sum_{i} N_i({\rm H~I})}
{ \Delta X}
\end{equation}

so that 

\begin{equation}\label{new_omega_eqn}
\Omega_{\rm DLA} = \frac{H_0 \mu m_H}{c \rho_{crit}} 
\frac{ \sum_{i} N_i({\rm H~I})}{ \Delta X}  
\end{equation}

from which the mass density of neutral gas traced by DLAs is
obtained by direct summation of the individual 
values of neutral hydrogen column density.  The redshift
path, $\Delta X$, which takes into account co-moving
distances, is given (in our adopted cosmology) by

\begin{equation}\label{delta_x}
\Delta X = \sum_{i} \frac{2}{3}[(1+z_{max,i})^{\frac{3}{2}} - 1] -
\frac{2}{3}[(1+z_{min,i})^{\frac{3}{2}} - 1] 
\end{equation}

\begin{figure}
\centerline{\resizebox{8cm}{!}
{\includegraphics{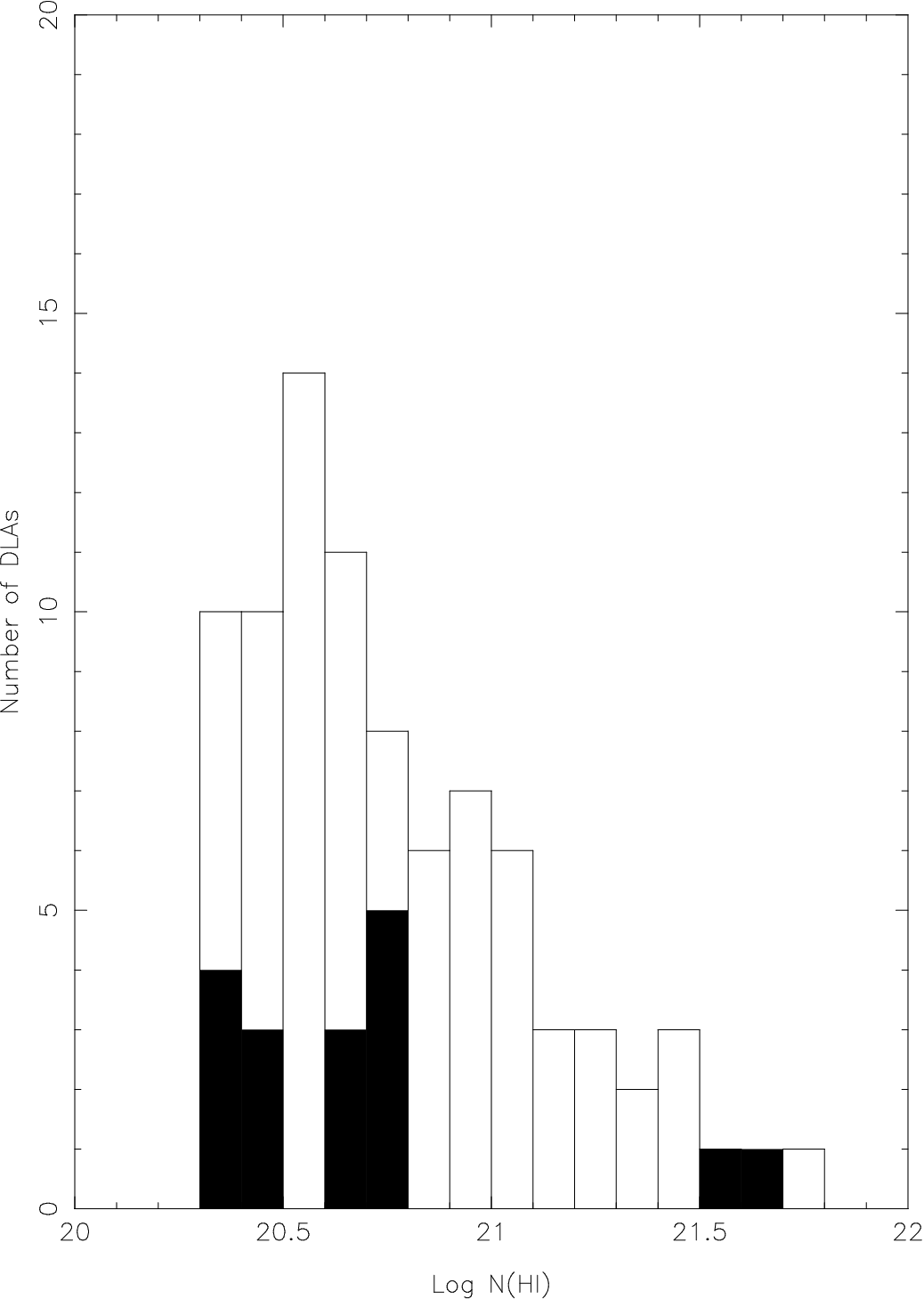}}}
\caption{\label{hi_dist} Distribution of H~I column
densities in the
large sample of Storrie-Lombardi \& Wolfe (2000) and in
the CORALS survey (shaded
histogram).  Due to the relatively small number of QSOs, and therefore
DLAs, included in CORALS, we do not fully sample the \nhi\ distribution.
}
\end{figure}

From eqs. (7) and (8) with the values of $N$(H~I) listed in
Table 3 we deduce $\log \Omega_{\rm DLA} h =
-2.59^{+0.17}_{-0.24}$ (with errors calculated as described by
Storrie-Lombardi et al. 1996).  As can be seen from 
Figure \ref{omega}, we again find that the CORALS value of
$\Omega_{\rm DLA}$ is in good agreement with previous
determinations. Thus, the most straight-forward conclusion
from the CORALS survey is that existing magnitude limited
samples of QSOs do not $seriously$ underestimate the number of DLAs, nor
their overall mass content; by and
large, they seem to provide a fair census of neutral gas at
high redshift. On the other hand, given the current
statistical uncertainties, the data still admit a moderate
degree of dust bias. The $1 \sigma$ limits on both $n(z)$
and $\Omega_{\rm DLA}$ include the possibility that analyses
based on optically selected samples may have underestimated both
quantities by a factor of $\sim 2$\,.
This possibility is further highlighted when we consider the
distribution of values of $N$(H~I) in the CORALS survey
(Figure \ref{hi_dist}). Given the small size of our survey,
we may well be missing DLAs at the high
column density end of the distribution simply through small
number statistics. We return to this point in section 6.3
below.  We note that although it is impossible to constrain the
H~I column density distribution for the CORALS sample, we do
detect 2 relatively rare high $N$(H~I) systems.  Therefore,
although we do not find evidence for a previously excluded population
of very high column density absorbers, with improved statistics
it would be of great interest to re-assess the $N$(H~I)
distribution function.

We also consider another effect. As explained in
section 2, many previous surveys have included in their
statistical analyses {\it candidate} DLAs, identified on the
basis of the equivalent width of the \lya\ line rather than
by profile fitting to a damped profile. It is worthwhile
examining the overestimate of $\Omega_{\rm DLA}$ which may
result from this approximation. For the 17 DLAs for which we
have obtained our own spectra, we measure the equivalent
width and compare the implied column density to that
determined by fitting the \lya\ line.  In most cases we find
the two techniques to be in very good agreement, certainly
within the errors associated with each method.  There are
only three exceptions (the DLAs in B1055$-$301, B1251$-$407a
and B2314$-$409) where the equivalent width determination
leads to a much higher \nhi\ than the line fit.  Inspection
of Figure \ref{all_DLAs} shows that this is due to extended
absorption around the DLA.  Although fits of these DLAs were
not straight-forward, this process was facilitated by higher
spectral resolution and coverage of metal lines which
provide additional guidance in the shape of the wings and
$z_{\rm abs}$.  Had we used the values of $N$(H~I) deduced from
the equivalent widths for the entire CORALS sample, we would have
over-estimated $\Omega_{\rm DLA}$ by 20\%\,. This discrepancy
would have been further increased if extended blends of
lines which do {\it not} include a DLA, such as those
present in the spectrum of B1251$-$407, were mistakenly included.
Nevertheless, the over-estimate is not large and, given the
increasing body of accurate measurements of $N$(H~I) in
DLAs, we think it very unlikely that this effect could be
masking a higher degree of dust bias than that indicated by
inspection of Figure 3.

Finally, for completeness, we calculate $\Omega_{\rm
DLA}$ in the redshift range $3.5 < z_{\rm abs} < 4.0$ where
we detect two DLAs despite the fact that with CORALS we only
sample a total interval $\Delta z=1.26$\,. The error bars
are naturally very large, but all the same it is intriguing
that $\log \Omega_{\rm DLA} h$ seems to remain high at $
-2.37^{+0.24}_{-0.59}$ in contrast with the slight down-turn
suggested by the work of Storrie-Lombardi \& Wolfe (2000)
and P\'{e}roux et al. (2001b). It will be very interesting to
see how better statistics will impact upon the CORALS value
of $\Omega_{\rm DLA}$ at the highest redshifts since 
the present determination would suggest an
increase in the importance of dust bias with increasing redshift.

\subsection{DLA Statistics as a Function of $B$-band Magnitude}\label{bmag_sec}

\begin{figure}
\centerline{\resizebox{11cm}{!}
{\includegraphics{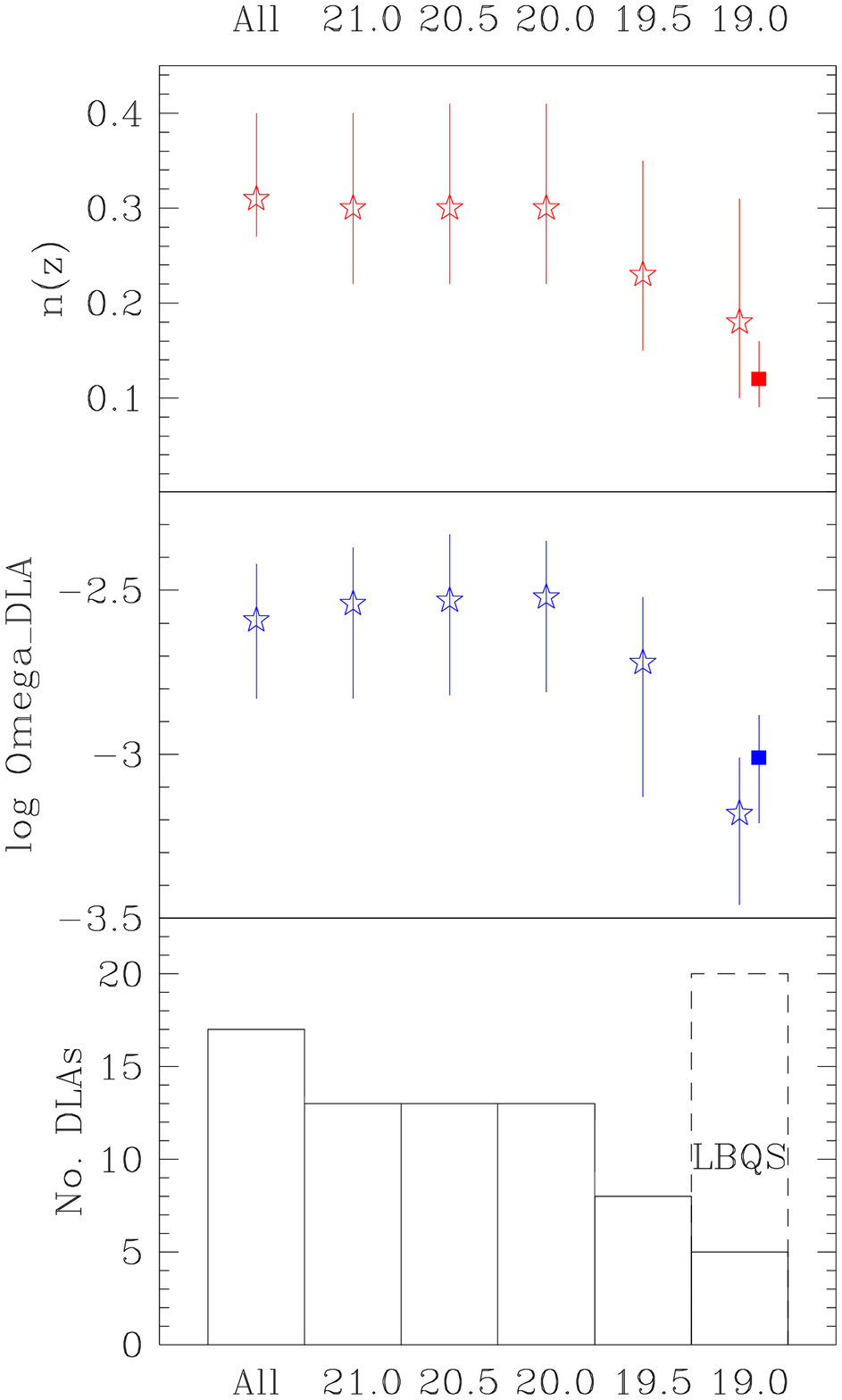}}}

\caption{\label{nhisto} DLA statistics as a function of
magnitude.  The quantities plotted are for the cumulative
values in the bins $\leq$ the $B$-band magnitude on the
$x$-axis. Bottom panel: Cumulative histogram of number of
DLAs. The dashed area shows the number of LBQS DLAs used to
determine improved statistics towards bright QSOs.  Middle
and top panels: stars show the values for CORALS DLAs,
whilst the solid square shows the same quantity for the LBQS
sample ($B \simlt 19$); this point has been offset on the
$x$-axis for clarity).  
}

\end{figure}

In their preliminary analysis of this sample, Ellison et al.
(2000) found tentative evidence that $\Omega_{\rm DLA}$ was
higher towards fainter QSOs, consistent with the effect
expected from a dust bias.    We re-examine this point in
Figure \ref{nhisto}, which shows cumulative statistics for
the CORALS DLAs as a function of the $B$-band magnitude of
the background QSOs. Since there are relatively few bright
QSOs in our sample, we also show the statistics for the LBQS
DLA survey (Wolfe et al. 1995) which has a limit $B \simlt
19$, using the column densities reported by Wolfe et al.
(1995) and Storrie-Lombardi \& Wolfe (2000). We confirm the
initial conclusion by Ellison et al. (2000) that
$\Omega_{\rm DLA}$ increases as fainter QSOs are observed, but
stabilises at $B \simeq 20.0$. Thus, we do {\it not} find a
population of high $N$(H~I) DLAs which is only revealed when
faint ($B \simgt 20$) QSOs are observed.
However, closer inspection of the data emphasises the need to extend
our survey in order to fully sample the column density 
distribution function, particularly at the high column
density end. The increase in $\Omega_{\rm DLA}$ between the $B
\leq 19.0$ and $B \leq 19.5$ bins is almost entirely due to
a single DLA with \nhi = $3.5 \times 10^{21}$ \cm.
Similarly, the increase between the $B \leq 19.5$ and $B
\leq 20.0$ bins is caused by a single DLA with \nhi = $4.5 \times
10^{21}$ \cm.  Between them, these two  systems account for
over half of the neutral gas in the entire DLA sample.  A K-S
test that compares the distribution of \nhi\ among DLAs
towards $B < 20$ QSOs with those from the large sample of
Storrie-Lombardi \& Wolfe (2000) provides inconclusive results, i.e.
that the samples are indistinguishable at only the 1$\sigma$ level.
Taken as a whole, the CORALS DLAs are \textit{inconsistent} (again only at the
1$\sigma$ level) with the \nhi\ distribution of 
Storrie-Lombardi \&  Wolfe (2000).  As found by Ellison et al. (2000)
for a sub-sample of the CORALS sample, we confirm that for
the complete sample that there is no strong correlation between
magnitude and redshift out to $z \sim 3.5$. Therefore, these
trends are not likely to be associated with the evolution
(apparent or real) of the properties of the QSOs themselves.

In addition to the cumulative values of $n(z)$ shown in
Figure \ref{nhisto}, we calculate the number density of
DLAs towards QSOs with $B \geq 20$ and $B < 20$, and find
$n(z) = 0.38^{+0.20}_{-0.14}$ (at $\langle z_{\rm abs}
\rangle = 2.54$) and $0.27^{+0.11}_{-0.08}$ (at $\langle
z_{\rm abs} \rangle = 2.25$)  respectively.  For the $B <
20$ subset, this value is consistent with the number density
found by Storrie-Lombardi \& Wolfe (2000).  Again, we see
that there is an excess of DLAs in faint QSOs, but only at
the $\sim 1 \sigma$ significance level.

\section{Summary and Discussion}

We have presented the first results from the CORALS survey
for DLAs in a radio-selected sample of QSOs.  The sample
consists of 66 $z_{\rm em} \geq 2.2$ QSOs, 58 of which have
been observed by us using the ESO 3.6~m, the AAT and the VLT
facilities, while the remaining eight were culled from the
literature. All the new spectra are presented in Figure 1. A
total of 22 DLAs has been identified; 19 of which are at
intervening redshifts.

We find that the comoving mass density of neutral gas implied by these
DLAs is $\log \Omega_{\rm DLA} h = -2.59^{+0.17}_{-0.24}$ at a
mean $\langle z_{\rm abs} \rangle = 2.37$, in good agreement
with previous surveys.  Similarly, the number density of
DLAs per unit redshift in radio-selected QSOs, $n(z)
=0.31^{+0.09}_{-0.08}$, is only $\sim 1 \sigma$ higher than
that in optical, magnitude-limited samples. Within our own
sample we also find that $n(z)$ is higher, but again by only
$\sim 1 \sigma$, in QSOs with $B > 20$, compared with
sightlines towards brighter quasars. These results indicate
that at redshifts $z = 2 - 3.5$ DLA surveys using optically
selected QSOs probably underestimate the number of DLAs, and
the gas mass they trace, by no more than a factor of about
2, in broad agreement with the predictions by  Pei \& Fall
(1995).  In particular, we have not uncovered a population
of high column density $N$(H~I)$ > 10^{21}$\,cm$^{-2}$
absorbers which had been missed in previous searches limited
to QSOs brighter than $B \simeq 20$.

These conclusions are somewhat tentative because of
the small size of the CORALS sample. Our value of
$\Omega_{\rm DLA}$ is dominated by two very high column
density systems, both of which occur in moderately bright
($B = 19.5, 20$) QSOs, and the column density
distribution function of DLAs is clearly not well sampled
with the relatively small number of QSOs in our survey.

The next important step in this work is to determine the
metallicities and dust content of CORALS DLAs.  The possibility
of a dust bias in DLA selection has previously been appealed to
in order to explain the lack of high column density, metal-rich
absorbers (e.g. Prantzos \& Boissier 1999). This explanation now
seems less likely, regardless of whether or not CORALS DLAs prove
to be more metal-rich, simply because high column density DLAs do
not appear to be significantly more common in fainter QSOs, at
least within the statistical limitations of our survey.
Nonetheless, determining the metallicities of the new DLAs
discovered here remains an important goal, because it will allow
us to assess whether the low element abundances found so far are
indeed typical of the full DLA population.

What could then be the reason for the observed dearth of high
column density, metal-rich DLAs?  We consider it unlikely that
gravitational lensing may be the answer. In principle one may
conjecture that close alignment of QSOs with foreground galaxies
may produce a tendency for such sightlines to be deflected away
from the inner regions of galaxies, where interstellar clouds
with high $N$(H~I) and high $Z$ may be preferentially
intercepted. However, to date no statistical evidence for lensing
of QSOs by DLAs has been found (Le Brun et al. 2000) and, in any
case, lensing would be most effective at significantly lower
redshifts than those considered here (Smette, Claeskens \&
Surdej 1997).

A more plausible explanation is that there is simply a
cross-sectional bias against detecting DLAs in sightlines that
pass through the centres of galaxies. Observationally, one could
argue that high $N$(H~I), high $Z$ DLAs have already been found
in the Lyman break galaxies (e.g. Pettini et al. 2000) which
indeed have typical linear sizes one order of magnitude smaller
than the impact parameters of most DLAs (Giavalisco, Steidel, \&
Macchetto 1996; Calzetti \& Giavalisco 2000; see also Figure 1 of
Pettini 2001). Theoretical studies of DLAs also support this
interpretation. For example, the models of Mathlin et al. (2001),
who simulated DLA surveys by sampling model galaxies at random
impact parameters, predict that the locus of high column density
and metal rich absorbers {\it should} be populated, but that DLAs
with these properties are intrinsically rare due to the small
cross-sectional area presented by the inner galactic regions
where they are found.  

As already emphasised, the CORALS data set is too small to sample
properly the column density distribution function and a
considerably larger survey is required in order to provide the
statistical coverage that will determine the true incidence of
\nhi $> 10^{21}$ \cm\ absorbers. A statistically larger survey will
not only improve our determinations of $\Omega_{\rm DLA}$ and
$n(z)$, but will also make it possible to investigate the
evolution (or lack thereof) these quantities with redshift. It will be
particularly interesting to examine the possibility of higher
dust bias at larger redshifts as suggested by the data in Figure 3; 
such data will offer an insight into the evolution
of dust at early epochs. One promising prospect for extending the
work presented here is the FIRST QSO survey (Gregg et al. 1996;
White et al. 2000). Although spectral follow-up has so far been
limited to bright $R < 19$ targets, future follow-up of FIRST
sources at fainter optical magnitudes would provide an excellent
complement to the CORALS survey.

\begin{acknowledgements}

It is a pleasure to acknowledge the consistent
support of this project by the ESO and AAT Time Assignment Panels
and the professional and efficient help of the telescope staff at the
AAT, ESO 3.6-m and VLT.  In particular, we are grateful to the
ESO Paranal science operations staff for their expert execution
of our service observations.  We thank Mauro Giavalisco and Lisa 
Storrie-Lombardi for obtaining spectra of two of our targets and Joop
Schaye for useful comments on an earlier draft of this paper.

\end{acknowledgements}

\end{document}